\documentclass[a4paper,11pt]{article}
\usepackage{jheppub} 
\usepackage{lineno}
\usepackage{slashed}
\usepackage[normalem]{ulem}

\title{\boldmath A thermofield-double model of Uhlmann's anholonomy}

\author{Péter Lévay$^1$ and Csaba Velich$^2$}

\affiliation{
$^{1}$ MTA-BME Quantum Dynamics and Correlations Research Group\\
Eötvös Loránd Research Network (ELKH)\\
Budapest University of Technology and Economics
\\M\H uegyetem rkp. 3., H-1111 Budapest, Hungary}
\affiliation{$^{2}$
Budapest University of Technology and Economics
\\M\H uegyetem rkp. 3., H-1111 Budapest, Hungary}

\emailAdd{levay.peter@ttk.bme.hu,csabusz.velich@gmail.com}

\abstract
{A simple parametrized family of quantum systems consisting of two entangled subsystems, dubbed left and right ones, both of them featuring $N$ qubits is considered in the thermofield double formalism.
We assume that the system evolves in a purely geometric manner based on the parallel transport condition due to Uhlmann.
We explore the different interpretations of this evolution relative to observers either coupled to  the left or to the right subsystems. The Uhlmann condition breaks the symmetry between left and right by regarding one of the two possible sets of local unitary operations as gauge degrees of freedom.
Then gauging the right side we show that the geometric evolution on the left manifests itself via certain local operations reminiscent of {\it non-unitary} filtering measurements. On the other hand on the right the basic evolutionary steps are organized into a sequence of {\it unitary} operations of a holonomic quantum computation. 
We
calculate the Uhlmann connection governing the transport for our model which turns out to be related to higher dimensional instantons. 
Then we evaluate the
anholonomy of the connection for geodesic triangles with geodesic segments defined with respect to the
Bures metric.
By analysing the explicit form of
the local filtering measurements showing up on the left side we realize that they are also optimal measurements for distinguishing two given mixed states in the statistical sense.
We also point out that by conducting an
interference experiment on the right side
one can observe the physical effects of the anholonomic quantum computation.
We demonstrate this by calculating explicit examples for phase shifts and visibility patterns arising in such interference experiments. Finally a sequence of geodesic triangles producing the iSWAP gate via anholonomy needed for computational universality is presented.}

\begin{document}
\maketitle
\flushbottom

\section{Introduction}

In this paper, we are studying a simple model system $\mathcal S$ consisting of two entangled subsystems. They will be called the $\ell$-subsystem (left) and the $r$-subsystem (right). Both of them will be consisting of $N$ qubits.
We are imagining $\mathcal S$ as  an entangled $2N$-qubit system that interacts with an environment $\mathcal E$ characterized by a set of externally driven parameters comprising a manifold.
Moreover, ${\mathcal S}$ will be regarded as a quantum
system, however ${\mathcal E}$ will be regarded classical.

At first sight this seems to be the usual setup of the Berry's Phase (BP) scenario\cite{Berry}.  There, a set of externally driven parameters gives rise to a parametrized family of quantum systems. The extra twist of our treatise will be that now the parametrized system features a family of {\it entangled states} (coupling the left and right subsystems) that are also of the thermofield double (TFD)form\cite{Winnig,Witten}.
Just like in the BP setup the  interacting  ${\mathcal S}+{\mathcal E}$ system evolves according to the time dependent Schrödinger equation.
In the BP case when the parameters change adiabatically the Scrödinger equation takes eigenstates to eigenstates of the instantaneous Hamiltonian. In this case it is well-known that there is a purely geometric contribution of the quantum evolution\cite{Berry}.
However, unlike in the BP scenario in this paper we will not be interested in the explicit form of the dynamics governed by a time dependent Hamiltonian which is driving the full ${\mathcal S}+{\mathcal E}$ system.
Rather we will be interested merely in the purely geometric part of the evolution of the states of ${\mathcal S}$.
The part that we are interested in is the one which is governed by a parallel transport condition demanded for entangled states.
We are simply assuming that the interaction in ${\mathcal S}+{\mathcal E}$ is somehow ensuring that such a condition holds.
In the BP scenario it is the quantum adiabatic theorem, that ensures this condition\cite{Berry}.
In this paper we are not attempting to reveal the physical mechanism which implements such a condition.
The paper is rather about exploring the physical consequences of such a condition.

\begin{figure}[!h]
    \centering\includegraphics[width=0.5\textwidth]{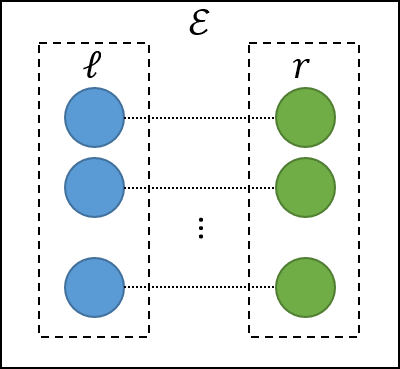}
    \caption{Illustration of the basic setup of this paper. An entangled system ${\mathcal S}={\ell }+{r}$} (consisting of the {\it left} (${\ell }$) subsystem and
    the ({\it r}) subsystem)
    is driven by another one $\mathcal E$ regarded as an environment. $\mathcal S$ is quantum and $\mathcal E$ is classical.
    Both $\ell$ and $r$ are consisiting of $N$-qubits. $\mathcal S$ is characterized by an entangled state.
    The simplest state of that kind is given by the vacuum state of Eq.(\ref{vacuum}) consisting of $N$ Bell pairs. The entanglement between left and right is indicated by dotted lines between the circles on the left and on the right. However, no connection is indicated between the blue circles on the left and the green ones on the right. This illustrates the fact that in the left and right reduced density matrices one has no entanglement or at most entanglement of bound type. This property is called quasi-classicality in the text.
    \label{fig:entang}
\end{figure}

In any case it is known that such a part of the evolution can be defined, without any recourse to a Hamiltonian, and in a purely geometric manner. Moreover, the resulting evolution of entangled states is then in general a non-unitary one\cite{Alsing}.
The underlying idea of such an evolution is based on the  possibility of relating two entangled states via the so called Uhlmann parallelity condition\cite{Uhlmann2,Uhlmann3,Uhlmann}.
Physically Uhlmann's parallelity is obtained by the extremization
of the interference between entangled states
with respect to either left or right local unitary operations.
This parallelity condition 
then yields parallel transport along geodesic segments in the space of
reduced density matrices of our entangled states with respect to the Bures metric\cite{Bures}.
The transport in turn gives rise
to an anholonomy effect first suggested by Uhlmann\cite{Uhlmann2,Uhlmann3,Uhlmann} as the mixed state generalization of Berry's Phase.

The model we consider will be characterized by a parametrized family of $2N$-qubit TFD states with both reduced density matrices providing a very simple $N$-qubit generalization of the usual Bloch ball model of a single qubit.
The reduced density matrices for our entangled states then will be of the thermal state form with one of the parameters $\beta$ corresponding to inverse temperature. While the angular parameters of this Bloch-sphere-like parametrization are collected into a unit vector ${\bf n}\in S^{2N}$.

Then in our model the entangled states will be parametrized by the set of coordinates 
$(\chi,\beta,{\bf n})$ where $0\leq\chi\leq 2\pi$, $0\leq \beta <\infty$.
By construction the extra parameter $\chi$ is featuring the entangled states but the marginals are independent of it. Occasionally we will use another parametrization of the form
$(\tau,r,{\bf n})$ where $-\infty<\tau<\infty$ and $0\leq r\leq \infty$.  Here $r$ can be regarded as a radial coordinate.
Some readers may find the (\ref{coord12}) coordinate transformations connecting these coordinates familiar from conformal field theory\cite{Perlmutter}.
Indeed, it is the Wick rotated version of the transformation used to transform the coordinates of the right Rindler wedge to the interior of a causal diamond\cite{Myers}.
In this respect one can regard the classical environment ${\mathcal E}$ as a manifold with  $(\tau,{\bf r})$ (with ${\bf r}=r{\bf n}$) being space time coordinates where $\tau$ is playing the role of Euclidean
time. This analogy will be particularly illuminating when we later realize that Uhlmann connection for our model is naturally related to
higher dimensional generalizations of
instantons\cite{Belavin,Zalanek}, objects which indeed live in a Wick rotated spacetime.

Apart from our reduced density matrices being of thermal state form they are also having another peculiar property. Namely that from the individual density matrices of the left and right subsystems no pure entanglement can ever be extracted. In this sense although the total system of left and right is quantum both left and right taken in itself is classical. 
This means that
for our model system the reduced density matrices of left and right have positive partial transposed. Hence their individual states can have at most bound entanglement\cite{Horo1}. This property of our subsystems  amenable for descriptions in terms of such density matrices will be dubbed "quasi-classicality".
The word quasi here refers to the fact that although it is known that no nonlocality associated with pure state entanglement can be distilled from them, but  such bound states still in principle can violate  a Bell's type inequality\cite{Vertesi}, hence they can still imply {\it some} nonlocality of other type.

Since our setup can be formulated in the thermofield double formalism\cite{Winnig,Witten}  naively one can envisage our system as an elementary model of a toy universe  with a "horizon" having the meaning here as the imaginary division line between left and right. 
Since our basic setup is rooted in this
framework one can also think of it
in terms of the ER=EPR setup\cite{EREPR}. In this context one can also imagine left and right as two sides of a situation reminiscent of the Einsten-Rosen (ER) bridge or the left and right sides of a Rindler wedge\cite{WittenAPS}. Indeed, one can think of the left system as a one which is although classically (e.g. spatially) separated,  but quantally it is entangled with its right partner.
Then the property of the quasi-classicality of bound mixed states for our subsystems on both sides mimics properties of dual spacetime regions where semiclassical methods are eligible.

This observation connects our work to the recent trend of investigating simple model systems in holographic situations connected to the AdS/CFT correspondence\cite{Maldacena}.
Indeed, since the work of Verlinde\cite{HVerlinde} it is obvious that in order to get insight into certain issues concerning quantum gravity it is rewarding to study simple quantum-mechanical examples of wormhole-like situations.
In particular it was shown that anholonomy effects connected to geometric phases distinguish between entangled states in wormhole quantum mechanics\cite{JErdmenger1,JErdmenger2}.
In this context a simple system of two entangled qubits has been used\cite{JErdmenger1}  to gain insight into the local time evolution properties of {\it one of the two} throats of a wormhole.
This situation is analogous to the one studied here where different types evolutions are identified and analysed for our subsystems.

More precisely our aim here is to explore how observers in contact with the left and right subsystems can interpret an overall geometric evolution of the entangled state of $\mathcal S$. As we will see the geometric evolution on the left manifests itself via certain local operations reminiscent of {\it non-unitary} filtering measurements. However, such an evolution has a wildly different meaning on the right. There the basic evolutionary steps are organized into a sequence of {\it unitary} operations. Moreover, it turns out that such a sequence is a quantum computation with its gates performed in a curious reversed manner. So in this toy model due to entanglement filtering measurements are interpreted as quantum computational steps in the right subsystem
"beyond the horizon".

Note in this respect that the asymmetry between left and right is arising from the existence of two possible choices for implementing the Uhlmann parallelity condition.
This ambiguity is connected to whether we regard the left or the right local unitary transformations as  gauge degrees of freedom.
In this paper we opted for the choice of gauging the degrees of freedom associated to the {\it right} unitaries
\footnote{For our choice see Eq.(\ref{Uparallel}). For the other one exchange the location of the $\ast$ referring to taking the adjoint.}.
Notice also that in our model left and right subsystems are reminiscent of the left and right Rindler wedges.
In this case it is known that the modular flow is purely geometric and is associated with the flow lines of a Lorentz boost\cite{WittenAPS}. Moreover, in this case left and right parts of the flow then are reversed with respect to one another.
The reverse order of operations on left and right
can then be regarded as an implementation of the directions of modular time in our discretized model of the Uhlmann setup.

In order to be able to study anholonomy effects the corresponding manipulations should be cyclic in nature. This means that the operators on the left should be implementing a sequence of filtering measurements that take the starting marginal back to its initial value.
Hence we will have a cycle in the space of reduced density matrices for the left subsystem.
On the other hand in the right subsystem this sequence will have the dual interpretation as a holonomic quantum computation\cite{Paolo}.
We again emphasize that this holonomic computation is not the usual one based on the non-Abelian Berry's phase scenario\cite{Berry, Wilczek} for pure states but rather the Uhlmann's one for mixed states\cite{Uhlmann2,Uhlmann}. 

Our elementary calculations can hopefully be considered as an appetiser for studying more realistic scenarios showing up within the context of holographic dualities.
Hence here we will be content with a detailed analysis of our very explicit toy example where  the basic structures are simple and they can be handled by analytical calculations.
In particular we will give the explicit form of the unitary operations providing the basic quantum gates featuring a quantum computation showing up in the right subsystem. These gates are built up from sequences of Thomas rotations familiar from special relativity.
Such rotations have already been studied in the mixed state context by one of us\cite{LPThomas}.
We also elaborate on the underlying geometric structures featuring these anholonomy effects.
Especially we calculate the gauge-field which is a pull-back of Uhlmann's connection governing the parallel transport.
It turns out that it is just a suitable restriction of the $SU(2^{N-1})$ gauge-fields responsible for higher dimensional monopoles\cite{Zalanek}, objects that has already been used in the theory of spontaneous compactification. For the $N=2$ case
the connection is related to the famous one of the $SU(2)$ Yang-Mills instanton\cite{Belavin,Yang,Rudolph,LPThomas}.
Next we will show that by conducting an interference experiment on the right subsystem, discussed in detail in the literature\cite{Sjöquist1}, we can observe the effects of the anholonomic quantum computation.

Finally we stress that the term quasi-classicality of our toy model is meant here that the local operations performed on the subsystems will be creating for the subsystems  entanglement of at most bound type.
However, this limitation is not fully exploited here. In our paper it is merely connected to the simplicity of our model which enables us to follow via explicit calculations how the results of filtering measurements are reinterpreted as quantum computational steps accross the "horizon". The full elaboration of the connection between quasi-classicality as defined here and Uhlmann anholonomy deserves further scrutiny.
The reader might also recognize that the duality of different interpretations of Uhlmann's anholonomy of an entangled state, viewed from the left or right perspective, is in accord with the relative state interpretation of an entangled state of  Everett\cite{Everett}.
Notice also that Uhlmann's anholonomy for mixed states and its physical implications have already made their debut to the holographic literature\cite{Kirklin1,Kirklin2}.
Further elaborations on connections with these results and our observations are worth pursuing and are planned in a future work.

The organization of this paper is as follows.
In Section 3. we start by summarizing some background material on invertible density matrices and their purifications. In Sections 4. and 5. we present our simple class of density matrices and their purifications. Different parametrizations are also considered here. Uhlmann's anholonomy and Uhlmann's connection is introduced in Sections 6. and 7.  In the latter section we calculate this connection for our model. The physical meaning of our connection is clarified in Section 8. Next in Section 9. we study the anholonomy effect. Here we also establish its intimate connection to Thomas rotations.
In Section 10. we calculate the anholonomy for geodesic triangles with geodesic segments defined with respect to the Bures metric.

Then from Section 11. onwards we turn to explorations on the physical meaning of Uhlmann's anholonomy in our TFD context.
First we discover that from the perspective of the left subsystem the anholonomy of a cycle can be cast in the form of a non-Abelian version of a sequence of filtering measurements. Alternatively one can regard these local filtering measurements as optimal measurements for distinguishing two given full rank mixed states in the statistical sense. This means that we need to make distinction between the two probability distributions associated with the two mixed states via applying sets of positive operator valued measures (POVM).
In our model setup we confirm a known result\cite{Ericsson,Fuchs}
claiming that the best measurement is provided by projectors associated to the geodesic (local filtering) operator.
In Section 12. by providing explicit formulas for the basic quantities of physical meaning we elucidate this result between geodesics and optimal measurements.

In Section 13. we turn to the right subsystem. 
Here we illustrate how the geometric evolution interpreted as local filtering measurements on the left can be reinterpreted as a quantum computation by observers on the right. We conclude that by conducting interference experiments of the type proposed in Ref.\cite{Sjöquist1} it is possible to extract physical information concerning this computation.
This is done by coupling an extra qubit to the right subsystem, a qubit that represents the two possibilities for paths of particles in a Mach-Zehnder interferometer.
Then we notice that in our setup unlike the left marginal,  the right marginal is changing due to Uhlmann anholonomy.
Then in this interferometric picture we can regard our right $N$-qubit subsystem as a system representing internal degrees of freedom for the particles subject to the two paths in the interferometer.
Then our change of the right marginal  can be viewed as a change occurring within this internal space of the particles subject to the measurement implemented by the interferometric setup. 
Then after subjecting this system to the interferometer the net result is that the output density matrix of our right subsystem plus the extra qubit will be producing a phase shift and the visibility of the interference pattern. 
Both of these observable quantities are featuring the anholonomy matrix calculated in previous sections.
We conclude by producing several explicit examples for phase shifts and visibility patterns, and a path consisting of geodesic segments producing the iSWAP gate.
This constitutes an example for producing universal quantum gates for quantum computing via designing convenient loops featuring geodesic
segments.
Finally the conclusions and some comments are presented in Section 13.
Some technical details can be found in the subsections of Section 15. compiled into an Appendix.

\section{Density matrices and their purifications}

We start by introducing some mathematical structures and then immediately elaborate on their physical meaning.
Let us denote the space of $n\times n$ matrices by ${\mathbb M}(n)$. The subspace of Hermitian $n\times n$ matrices will be referred to as  ${\mathbb H}(n)$
and the subset of strictly positive Hermitian matrices  as ${\mathbb P}(n)$. 

In this paper we wish to study a physical model system represented by a special subset of 
${\mathbb P}(n)$ comprising
{\it invertible} density matrices. A typical element of that kind will be denoted by $\varrho$.
For such density matrices apart from $\varrho\in{\mathbb P}(n)$, we also have ${\rm Det}(\varrho)\neq 0$ and ${\rm Tr}(\varrho)=1$. The space of such full rank density matrices will be denoted by ${\mathcal D}\subset {\mathbb P}(n)$.

For a density matrix $\varrho\in{\mathcal D}$ the matrices $W\in GL(n):=GL(n,{\mathbb C})$ satisfy the constraints
$\varrho =WW^{\ast}$ and
$\langle W\vert W\rangle =\vert\vert W\vert\vert^2 =1$,
where
$\langle W_1\vert W_2\rangle:={\rm Tr}(W_1^{\ast}W_2)$
comprise the space of special purifications of $\varrho$.
Such a $W$ will represent an entangled state coupling two subsystems. 
Here by$\ast$ we refer to the adjoint.

Consider now the map
$\pi:GL(n)\to {\mathbb P}(n)$ defined by $\pi(W)=WW^{\ast}$.
$GL(n)$ is a Riemannian manifold with the metric induced by the Frobenius inner product 
$(W_1,W_2):={\rm Re} \langle W_1\vert W_2\rangle$. The group $U(n)$ acting on $GL(n)$ from the right as $W\mapsto WU$ with $U\in U(n)$ is a group of isometries for this metric\cite{Bhatia,Bhatia2}. Then the quotient space $GL(n)/U(n)$ is ${\mathbb P}(n)$. 
The map $\pi$ defines a trivial principal bundle.
Using the projection $\pi$ and the metric on $GL(n)$ one can induce a metric on the base space ${\mathbb P}(n)$ as well. This metric is called the Bures metric\cite{Bures}.
For $A,B\in{\mathbb P}(n)$, the associated Bures-Wasserstein distance\cite{Bhatia2} can be computed as 
\begin{equation}
d^2(A,B)={\rm Tr}(A)+{\rm Tr}(B)- 2{\rm Tr}(A^{1/2}BA^{1/2})^{1/2}    
\label{Wasserstein}
\end{equation}
It is also well-known that
\begin{equation}
d(A,B)=
\min_{U\in U(n)} \vert\vert A^{1/2}-B^{1/2}U\vert\vert  
\label{Buresertem}
\end{equation}
\noindent
where $\vert\vert A\vert\vert^2={\rm Tr}(A^{\ast}A)$.

Let us clarify the physical meaning of these abstract mathematical concepts.
A special purification\footnote{The constructions to be discussed here are motivated by the well-known thermofield double considerations\cite{Winnig,Witten}. In such cases both of the subsystems have the same dimension. Of course there can exist other purifications that give rise to the same reduced density matrices.} of an $n\times n$ density matrix $\varrho$ can be regarded as an entangled state  
$\vert\Psi\rangle\in {\mathcal H}_{\ell}\otimes {\mathcal H}_{r}$ where states of the subsystems are elements of {\it two} copies of $n$ dimensional Hilbert spaces. The corresponding subsystems will be called the left subsystem and the right subsystem.
Then a state of the composite system can be written in the form
\begin{equation}
\vert\Psi\rangle=    
\sum_{I=1}^n\sum_{J=1}^n W_{IJ}\vert I\rangle_{\ell}\otimes \vert J\rangle_{r}
\label{sokqubit}
\end{equation}
Here $W\in GL(n)$ and $\vert I\rangle_{\ell}$ and $\vert J\rangle_r$ are orthonormal basis vectors of the corresponding Hilbert spaces.
The constraint $\langle W\vert W\rangle =\vert\vert W\vert\vert^2 =1$ represents the normalization condition for $\vert \Psi\rangle$.

When we have operators $(A\otimes {\bf 1})({\bf 1}\otimes B)=(A\otimes B)$ acting on ${\mathcal H}_{\ell}\otimes{\mathcal H}_r$ we have a map 
\begin{equation}
    \vert\Psi\rangle \mapsto (A\otimes B)\vert \Psi\rangle
    \label{corr1}
\end{equation}
or alternatively in a matrix representation
\begin{equation}
W\mapsto A W B^T
\label{corr2}
\end{equation}
where $A,B\in {\mathbb M}(n)$ and  $B^T$ refers to the transposed of $B$. 

Such operators represent local manipulations to be performed on only one of the subsystems.
Here $A$ acts on the left subsystem and $B$ acts on the right one.
By tracing out the right or the left subsystem one can then form the reduced density matrices of the left and right subsystems
\begin{equation}\varrho_{\ell}=WW^{\ast},\qquad
\varrho_r=\overline{W^{\ast}W}
\label{firstex}
\end{equation}
In the following the density matrices for the left and right subsystems will be denoted by different letters namely
\begin{equation}
 \varrho:=\varrho_{\ell}\in{\mathcal D},\qquad
 \omega:=\varrho_{r}\in{\mathcal D}
\label{megallapodas}
\end{equation}

In the first expression of
(\ref{firstex}), after the identification $\varrho:=\varrho_{\ell}\in{\mathcal D}$, we recognize the defining equation of our $U(n)$ principal bundle $\pi: GL(n)\to {\mathbb P}(n)$. Here the $U(n)$ gauge degree of freedom amounts to transformations of the form $\vert\Psi\rangle \mapsto ({\bf 1}\otimes U)\vert\Psi\rangle$
with $U\in U(n)$ representing transformations  effected on the right subsystem. Alternatively we have the {\it right} action $W\mapsto WU^T$ leaving   
the state $\varrho$ of the observer in the left subsystem invariant.
Notice that here we opted for choosing the unitary degrees of the {\it right system} as gauge degrees of freedom. Of course we could have chosen the other way round.
Our choice boils down to the invariance of the {\it left} marginal $\varrho_{\ell}$ under the {\it right} action $W\mapsto WU^T$. We could have also chosen the possibility of gauging the left action however, it will makes no difference other than exchanging the role of the two subsystems.

Let us finally comment on the physical meaning of the Bures distance obtained from Eq.(\ref{Wasserstein})
by choosing $A=\varrho_1$ and $B=\varrho_2$ with both of them satisfying ${\rm Tr}(\varrho_j)=1, j=1,2$. 
Their purifications can be written in the polar decomposed form : $W_j=\varrho_j^{1/2}U_j$, with $\varrho^{1/2}_j\in {\mathbb P}(n)$ and $U_j\in U(n)$.
The special purification $W_{0j}:=\varrho_j^{1/2}$ will be called the {\it canonical purification} of $\varrho_j$.
For any $\varrho\in{\mathcal D}=GL(n)/U(n)$ the canonical purification $\varrho^{1/2}$ provides a global section of our principal bundle.

The canonical purification $W_{0j}\in{\mathbb P}(n)$ can be reinterpreted as  an entangled state
$\vert \varphi_j\rangle\in{\mathcal H}_{\ell}\otimes {\mathcal H}_r$ i.e.
\begin{equation}
    \varrho^{1/2}\in {\mathbb P}(n)\leftrightarrow
    \vert\varphi\rangle\in{\mathcal H}_{\ell}\otimes {\mathcal H}_r
    \label{correspondence}
\end{equation}
An explicit example of this correspondence will be given in the next section.
Using this one can understand the physical meaning of the (\ref{Buresertem}) Bures distance as follows.
The Bures distance squared $d^2(\varrho_1,\varrho_2)$ for two mixed states of the {\it left} subsystem is calculated via finding the purifying entangled state $\vert \psi_2\rangle =(I\otimes U)\vert\varphi_2\rangle$
of the combined left plus right system, such that its interference with the state $\vert\psi_1\rangle:=\vert\varphi_1\rangle$ is minimal.
By minimal interference we mean the quantity obtained after finding the minimal value
$\min_{U} \vert\vert \varphi_1-(I\otimes U)\varphi_2\vert\vert^2=2(1- \max_{U} {\rm Re}\langle \varphi_1\vert(I\otimes U)\varphi_2\rangle)$.

Notice that in our setup in order to define the distance between two states in the {\it left} subsystem a minimization in the space of unitaries acting on the {\it right} system has to be performed.
Moreover,
the physical origin of
this notion of "distance" rests in the phenomenon of interference between entangled states.

\section{A simple class of full rank density matrices}

In the previous section we introduced the space of full rank density matrices $\mathcal D\subset{\mathbb P}(n)$ and their  purifications.
In this paper however, we will consider only a convenient very simple  $N$-qubit subset ${\mathcal B}\subset {\mathcal D}$ useful for our purposes.
Hence for our model we have $n=2^N$.
Moreover, this subset ${\mathcal B}$ will be parametrized by the interior of the $2N$ dimensional unit ball. For $N=1$ we get back to the interior of the well-known Bloch-ball.

Explicitly let ${\mathbb B}^{2N}:=\{{\bf u}\in {\mathbb R}^{2N+1}\vert \vert {\bf u}\vert< 1\}$ and define
\begin{equation}
    {\mathcal B}:=\{\varrho\in{\mathcal D}\quad\vert \quad \varrho =
\left(I+\slashed{u}\right)/2^N,\quad {\bf u}\in
{\mathbb B}^{2N}\}
\label{Ballset}
\end{equation}
where
\begin{equation}
\slashed{u}:={\bf u}\boldsymbol{\Gamma}=u^{j}\Gamma_{j}=\delta_{jk}u^j{\Gamma^k},\qquad j=0,1,2,\dots, 2N
\label{density}
\end{equation}
$I$ is the $2^N\times 2^N$ identity matrix
and in the following summation for repeated indices is understood.
The $2^N\times 2^N$ gamma matrices satisfy
\begin{equation}
\{{\Gamma}_{j},{\Gamma}_{k}\}=2\delta_{jk}I
\label{anti}
\end{equation}
For possible conventions concerning gamma matrices used in this paper see Appendix 15.1.

Let us define the eigenvectors and eigenvalues of $\varrho$ as
\begin{equation}
\varrho\vert I\rangle=\lambda_I\vert I\rangle,\qquad I=1,2,\dots ,2^N    
\label{iii}
\end{equation}
Since ${\slashed{u}}^2=\vert{\bf u}\vert^2 I$ we have
\begin{equation}
\lambda_{\alpha}=\frac{1}{2^N}(1+\vert {\bf u}\vert)>0,\qquad
\lambda_{\bar{\alpha}}=\frac{1}{2^N}(1-\vert {\bf u}\vert)>0
\label{iiii}
\end{equation}
where $\alpha=1,2,\dots, 2^{N-1}$ and 
$\bar{\alpha}=1,2,\dots, 2^{N-1}$.
With this notation we use the split
$I=(\alpha,\bar{\alpha})$
hence we have two sets of $2^{N-1}$-fold degenerate eigenstates, namely
$\vert \alpha\rangle$ and $\vert\bar{\alpha}\rangle$.

We can check whether our $N$ qubit mixed state is entangled or not.
In order to check this one can consider an arbitrary bipartition and apply the transpose criterion of Peres\cite{Peres}. This is a necessary condition of separability. Notice then that since gamma matrices can be defined recursively (see Appendix 15.1) the relevant part of our density matrix subject to partial transpose is a linear combination of lower order gamma matrices. These are either symmetric or antisymmetric matrices. Then the effect of partial transposition can only boil down to a sign change of the relevant component of the vector ${\bf u}$ corresponding to the presence of antisymmetric submatrices in the relevant part of the tensor product. 
Now the (\ref{iiii})  eigenvalues of $\varrho$ belonging to our class are depending merely on the magnitude of ${\bf u}$. Then partial transpose cannot yield any negative eigenvalue.

Since the transpose criterion for $N=2$ is a sufficient and necessary criterion then in this case it means that $\varrho$ is separable. However, for $N\geq3$ this criterion is only necessary hence $\varrho$ can only contain at most bound entanglement\cite{Horo1}. 
This means that from this density matrix no pure entanglement can ever be extracted.
Note, however that although from these states no nonlocality associated with pure state entanglement can be distilled, but such states still can in principle violate Bell's type inequalities\cite{Vertesi}. In this respect our states taken from the class $\mathcal B$ can still imply {\it some} nonlocality of other type. 
In the rest of this paper we refer to this property of our class ${\mathcal B}$ as {\it quasi-classicality} of the corresponding $N$-qubit density matrices.

\section{Purifications of our class}

Now we would like to study purifications of a $\varrho\in{\mathcal B}$.
For a general $\varrho\in{\mathcal D}$
via polar decomposition one can define the set of purifications of $\varrho$ as
\begin{equation}
\mathcal{F}(\varrho):=\{ W\in GL(n)\vert\quad W=\varrho^{1/2}U,\quad U\in U(n)\}
\label{calf}
\end{equation}
Then in particular for our purifications coming from ${\mathcal B}$ we will use the notation
\begin{equation}
{\mathcal W}:=\{W\in {\mathcal F}(\varrho)\quad\vert \varrho\in{\mathcal B}\}.    
\end{equation}
where now $n=2^N$.

Let us now define the coordinates
\begin{equation}
{\bf r}=r{\bf n},\qquad \tau=\frac{\sin\chi}{\cosh\beta+\cos\chi}    
,\qquad r=\frac{\sinh\beta}{\cosh\beta+\cos\chi} 
\label{coord12}
\end{equation}
where $\vert{\bf n}\vert^2=1$ i.e. ${\bf n}\in S^{2N}$ and
\begin{equation}
    0\leq\chi\leq 2\pi,\qquad 0\leq \beta<\infty
\end{equation}
Then we have $r\geq 0$ showing that it can be used as a radial coordinate.
Let us also define the quantity
\begin{equation}
    \Omega:=(\cosh\beta+\cos\chi)^{-1}> 0
\end{equation}
Then we also have
\begin{equation}
    1+\tau^2+r^2=2\Omega\cosh\beta\qquad
    1-\tau^2-r^2=2\Omega\cos\chi
\end{equation}

Now we define our central quantity of interest. It is the following simple parametrized family of $2^N\times 2^N$ matrices
\begin{equation}
    W:=W(\tau,r,{\bf n})=\frac{1}{\sqrt{2^N(1+\tau^2+r^2)}}\left[(1+i\tau)I-r \slashed{\bf n}\right]
\label{purify}
\end{equation}
Soon we will see that
$W\in{\mathcal W}$ i.e. it can be regarded as a {\it special} family of purifications for our density matrices of the (\ref{density}) form.
At the same time, according to Eq.(\ref{sokqubit}), one can also regard $W$ as a matrix comprising the amplitudes of a {\it special} $2N$ qubit state where both the left and right system contains N qubits.  
In the following we will regard $W$ as the basic mathematical object defining a quantized physical system of $2N$ qubits.
On the other hand the coordinates $\tau$, $r$ and ${\bf n}$ or alternatively $\chi$, $\beta$ and ${\bf n}$ are external parameters, which are not quantized. In the appendix it is shown that this alternative parametrization describes the configuration space coinciding with the interior of the unit ball times a circle, i.e. ${\mathbb B}^{2N}\times S^1$.
Moreover, in the spirit of Eq.(\ref{correspondence}) an entangled state $\vert\psi\rangle$ answering $W$ should rather be denoted as $\vert \psi(\chi, \beta, {\bf n})\rangle$. Hence our state is depending on externally driven parameters. 
 
Some readers may find the (\ref{coord12}) coordinate transformations familiar from conformal field theory\cite{Perlmutter}.
Indeed, it is the Wick rotated version of the transformation used for transforming the coordinates of the right Rindler wedge to the interior of a causal diamond\cite{Myers}.
In this respect one can regard our set of parameters $(\tau,{\bf r})$ as space time coordinates with $\tau$ plying the role of Euclidean
time of some spacetime coordinate patch. This analogy will be particularly illuminating when we realize in Section 8. that Uhlmann connection for our model is naturally related to instantons, objects which live in Wick rotated spacetime.
Though suggestive, we are not elaborating on this analogy further here. We are simply using the pair $(\tau,{\bf r})$ as convenient and instructive parametrization for our purification.

Notice also that when elaborating on the meaning of $W$ it is also useful to regard 
\begin{equation}
H({\bf n}):=\slashed{\bf n}\in {\mathbb H}(n)
\label{hami}
\end{equation}
as a Hamiltonian parametrized by the points of the unit sphere $S^{2N}$ with its $2^{N-1}$-fold degenerate energy eigenvalues being  $E_{\pm}=\pm 1$.
First note that by diagonalizing (\ref{hami}) one can also check that 
\begin{equation}
    {\rm Det}W=\left(\frac{e^{i\chi}}{2^N\cosh\beta}\right)^{2^{N-1}}\neq 0
\label{asfollows}
\end{equation}
hence $W$ is invertible.
Then observe that
\begin{equation}
    \varrho=WW^{\ast}=\frac{1}{2^N}(I-\tanh\beta \slashed{\bf n})=\frac{1}{2^N\cosh\beta}e^{-\beta H({\bf n})}
=\frac{e^{-\beta H({\bf n})}}{{\rm Tr}(e^{-\beta H({\bf n})})}
\label{alter}
\end{equation}
i.e. the (\ref{Ballset}) invertible density matrix can also be written in the thermal state form with $\beta$ playing the role of the inverse temperature and we have ${\bf u}\equiv -\tanh\beta{\bf n}$.

In Appendix 15.2 we show that
$W$ of (\ref{purify}) can be written in the polar decomposed form
\begin{equation}
W={\varrho}^{1/2}U=
U{\varrho}^{1/2}
, \qquad U^{\dagger}=U^{-1}   
\label{polar1}
\end{equation}
where several forms for the unitary are also given.
A useful form of this decomposition to be used later is
featuring the quantities
\begin{equation}
    {\varrho}^{1/2}=\frac{I+\slashed{a}}{\sqrt{{2^N}(1+a^2)}},\qquad {\bf a}:=-\tanh\frac{\beta}{2}{\bf n}
\label{root}
\end{equation}
\begin{equation}
U=e^{i\chi/2}\sqrt{\frac{\cosh\beta +1}{\cosh\beta +\cos\chi}}\left[\cos\left(\frac{\chi}{2}\right) I +i\sin\left(\frac{\chi}{2}\right)\slashed{a}\right]    
\end{equation} 
The geometric meaning of the coordinate ${\bf a}$ will be clarified in Section 8 and in the Appendix. 

Obviously $W$ with the polar decomposition just discussed is only a special element of the space $\mathcal W$. For a generic element $w\in\mathcal W$ one has $w=\varrho^{1/2}{\mathcal U}$ where $\mathcal U$ is an arbitrary element of $U(2^N)$.
Moreover, later we learn that when making a further restriction to ${\mathcal U}\in Spin(2N+1)$ a suitable subset of ${\mathcal W}$ can be regarded as a principal bundle over ${\mathbb B}^{2N}$.

 By virtue of the (\ref{correspondence}) correspondence and  Eq.(\ref{para}) of the Appendix one can check that $\varrho^{1/2}$ can be cast in the thermal field double\cite{Winnig,Witten} (TFD) form where $\beta$ is playing the role of the inverse temperature.
To check this notice that in the notation of Eq.(\ref{sokqubit})
the TFD state is
defined as\cite{Witten}
\begin{equation}
\vert\varphi\rangle=\frac{1}{\sqrt{Z_N}} \sum_{I=1}^{2^N}e^{-\beta E_I/2}\vert I\rangle_{\ell}\otimes\vert \bar{I}\rangle_r,\qquad Z_N=\sum_{I=1}^{2^N}e^{-\beta E_I}   
\label{zefi}
\end{equation}
where the $E_I$ are the eigenvalues of some multiqubit Hamiltonian.
The name TFD originates from the observation that
after tracing out the right subsystem one ends up with the left reduced density matrix $\varrho:=\varrho_{\ell}$ of the 
\begin{equation}
\varrho=\frac{1}{Z_N}\sum_{I=1}^{2^N}e^{-\beta E_I}\vert I\rangle\langle I\vert
\label{zefi2}
\end{equation} thermal state form. For our model we have $Z_N=2^N\cosh\beta$.
Moreover, according to Eq.(\ref{correspondence})
we see that the entangled state (\ref{zefi}) corresponds to the canonical purification with $E_{\alpha}=+1$ and $E_{\bar{\alpha}}=-1$ and the eigenstates are $\vert \alpha\rangle$ and $\vert\bar{\alpha}\rangle$ answering the (\ref{iiii}) split $I=(\alpha,\bar{\alpha})$.

The TFD state $\vert\varphi\rangle\leftrightarrow \varrho^{1/2}$ is an entangled state describing the entanglement between the left and right $N$ qubit subsystems. The von Neumann entropy of this entangled state can be calculated from the formula
\begin{equation}
S=-{\rm Tr}(\varrho\log\varrho)=-\sum_I\lambda_I\log\lambda_I=\log\left(2^N\cosh\beta\right)-\beta    
\tanh\beta=\log Z_N+\beta\langle\slashed{n}\rangle_{\varrho}
\label{freeenergy}
\end{equation}
where we have used (\ref{iiii}).
Since $H=\slashed{n}$ the last term is the expectation value of the energy $E$ in the state $\varrho$. Then using $\beta=1/T$ and $F=-T\log Z$ for the free energy we see that Eq.(\ref{freeenergy}) is in accord with the formula $F=E-TS$ familiar from statistical physics.

\section{Uhlmann anholonomy}

Here we introduce the general concepts of Uhlmann anholonomy\cite{Uhlmann,Uhlmann2}. Later we will specify these ideas for our (\ref{Ballset}) special density matrices and their purifications discussed in Section 4.

For two matrices $A,B\in {\mathbb P}(n)$ we define their geometric mean as another matrix $A\sharp B\in{\mathbb P}(n)$ of the form\cite{Pusz}
\begin{equation}
A\sharp B:=A^{1/2}\left(A^{-1/2}BA^{-1/2}\right)^{1/2}A^{1/2}\in{\mathbb P}(n)
\label{geomean}
\end{equation}
It is known\cite{Bhatia} that $A\sharp B=B\sharp A$ 
and that
\begin{equation}
A\sharp B=A(A^{-1}B)^{1/2}=(AB^{-1})^{1/2}B    
\label{Bhatiaid}
\end{equation}
One can also prove that for $A,B\in {\mathbb P}(n)$ the following matrix is unitary, i.e.
\begin{equation}
U:=B^{-1/2}A^{-1/2}(A^{1/2}BA^{1/2})^{1/2}\in U(n)
\label{y}
\end{equation}
One then has
\begin{equation}
B^{1/2}UA^{-1/2}=A^{-1}\sharp B=B\sharp A^{-1}    
\label{imp1}
\end{equation}
If we restrict our attention to density matrices belonging to ${\mathcal D}$ then for $A=\varrho_1$ and $B=\varrho_2$ from (\ref{imp1}) we get
\begin{equation}
(\varrho_2\sharp\varrho_1^{-1})\varrho_1^{1/2}=\varrho_2^{1/2}U  
\label{extra}
\end{equation}
With $W_1=\varrho_1^{1/2}$ and $W_2=\varrho_2^{1/2}U$ 
one can also write this as $W_2=(\varrho_2\sharp\varrho_1^{-1})W_1$.
Since $\varrho_j=W_jW_j^{\ast}$ for $j=1,2$ from this it is easy to check that
\begin{equation}
    W_1^{\ast}W_2>0
    \label{Uparallel}
\end{equation}
Two purifications satisfying the above constraints are said to be
Uhlmann-parallel\cite{Uhlmann}.
Notice that due to positivity  
we also have $W_1^{\ast}W_2=W_2^{\ast}W_1$.
This parallelism defines a connection on our bundle $GL(n)$ (or $\mathcal W$). The explicit form of the Uhlmann connection for the bundle ${\mathcal W}$ will be studied in the next section.

For the time being let us elaborate on the Uhlmann anholonomy of this connection. Let us consider a third point $W_3$ 
in our bundle. 
Clearly it is parallel with respect to $W_2$ iff
$W_3=(\varrho_3\sharp\varrho_2^{-1})W_2$.
Continuing in this manner we obtain the following sequence
\begin{equation}
    W_{k+1}=
    (\varrho_{k+1}\sharp\varrho_{k}^{-1})
    (\varrho_k\sharp\varrho_{k-1}^{-1})
    \dots(\varrho_3\sharp\varrho_2^{-1})(\varrho_2\sharp\varrho_1^{-1})W_1
\label{one}
\nonumber
\end{equation}
If at the $k+1$-th step we choose a purification on the same fiber that we have started with, we have $W_{k+1}=W_1{\mathcal U}$
for some ${\mathcal U}\in U(n)$.

Let us now use Eq.(\ref{extra}) in the form
\begin{equation}
U_{ij}= \varrho_i^{-1/2}(\varrho_i\sharp\varrho_j^{-1})\varrho_j^{1/2}
\label{baromi}
\end{equation}
Then using this rule iteratively one obtains
\begin{equation}
W_{k+1}=W_1{\mathcal U}=W_1U_{k+1k}U_{kk-1}\cdots U_{32}U_{21}
\label{two}
\end{equation}
\begin{equation}
U_{ij}=\varrho_i^{-1/2}\varrho_j^{-1/2}(\varrho_j^{1/2}\varrho_i\varrho_j^{1/2})^{1/2}\in U(n)
\label{yij}
\end{equation}
Recalling that $W_1=\varrho_1^{1/2}$ we obtain our final result
\begin{equation}
(\varrho_{1}\sharp\varrho_{k}^{-1})
(\varrho_k\sharp\varrho_{k-1}^{-1})
    \dots(\varrho_3\sharp\varrho_2^{-1})(\varrho_2\sharp\varrho_1^{-1})W_1=
W_1
U_{1k}
U_{kk-1}U_{kk-1}\cdots U_{32}U_{21}
\label{nice}
\end{equation}
Hence from the mathematical point of view one can implement Uhlmann's parallel transport as follows. Let us act on the left subsystem of the entangled state corresponding to the starting purification $W_1$ by a sequence of certain ${\mathbb P}(n)$ elements. According to Eq.(\ref{nice}) it will be producing a sequence of $U(n)$ transformations on the right subsystem of $W_1$.
The collection $\mathcal U =U_{1k}U_{kk-1}\cdots U_{32}U_{21}\in U(n)$
is called the Uhlmann anholonomy of $W_1$.

Notice that by virtue of the identities
\begin{equation}
\varrho_i\sharp \varrho_j^{-1}=\varrho_i(\varrho_j\varrho_i)^{-1/2},\quad
\varrho_j\sharp \varrho_i^{-1}=(\varrho_j\varrho_i)^{1/2}\varrho_i^{-1}
\end{equation}
the first of which can be obtained from the first equality Eq.(\ref{Bhatiaid}) and the second of which from the second equality of Eq.(\ref{Bhatiaid}) one obtains using (\ref{baromi}) the result
\begin{equation}
    U_{ij}U_{ji}=I
\label{inverseunitary}
\end{equation}
Then by virtue of our basic (\ref{correspondence}) correspondence using Eqs. (\ref{corr1}), (\ref{corr2}), (\ref{nice}) and the unitarity of $U_{ij}$ one obtains the final result
\begin{equation}
(L_{1k}L_{kk-1}\cdots L_{32}L_{21}\otimes I)\vert\varphi_1\rangle=(I\otimes R_{12}R_{23}\cdots R_{k-1k}R_{k1})\vert\varphi_1\rangle    
\label{essence}
\end{equation}
where 
\begin{equation}
    L_{ij}:=\varrho_i\sharp\varrho_j^{-1}\in{\mathbb P}(n),\qquad R_{ij}=\overline{U_{ij}}\in U(n)
\label{geomean1}
\end{equation}
Hence by Uhlmann parallelism manipulations performed on the left system, forming a sequence
of Hermitian positive observables,
can dually be interpreted as manipulations on the right system forming a sequence of unitaries {\it in the reversed order}.

\section{Uhlmann connection}

For a parametrized family $w(t)$
of purifications ($\varrho(t)=w(t)w^{\ast}(t)$) one can 
decompose the tangent vector $\dot w(t)$ to horizontal ($\mathcal{G}$) and vertical ($\mathcal{A}$) components as follows
\begin{equation}
\dot{w}={\mathcal G}w+w{\mathcal A},\qquad {\mathcal G}^{\ast}={\mathcal G},\quad {\mathcal A}^{\ast}=-{\mathcal A}
\label{clear}
\end{equation}
Here the vertical direction corresponds to the anti-Hermitian generator ${\mathcal A}$ of the right $U(n)$ action. Hermiticity and left multiplication by ${\mathcal G}$ is needed to ensure that
\begin{equation}
\langle {\mathcal G}w\vert w{\mathcal A}\rangle+
\langle w{\mathcal A}\vert {\mathcal G}w\rangle =0
\end{equation}
i.e. the orthogonality of the horizontal and vertical parts with respect to the Frobenius inner product $(X,Y):={\rm Re}\langle X\vert Y\rangle$
on $GL(n)$.

Then
\begin{equation}
\vert\vert \dot w\vert\vert^2=
\vert\vert{\mathcal G}w\vert\vert^2+
\vert\vert w{\mathcal A}\vert\vert^2
\nonumber
\end{equation}
From here one can see that the special choice for horizontal vectors $\dot{W}$ having the "minimal length" property 
$\vert\vert \dot W\vert\vert^2=
\vert\vert {\mathcal G}W\vert\vert^2$
are also satisfying
\begin{equation}
\dot W={\mathcal G}W \leftrightarrow \dot{W}^{\ast}W=W^{\ast}\dot{W}    
\end{equation}
which is just the infinitesimal version of Eq.(\ref{Uparallel}).
By virtue of Eq.(\ref{clear})
in the nonparallel case we have\cite{Uhlmann,Rudolph,kinaiak}
\begin{equation}
\dot{w}^{\ast}w-w^{\ast}\dot{w}={\mathcal A}w^{\ast}w+w^{\ast}w{\mathcal A}
\end{equation}
which serves as an equation for determining ${\mathcal A}(t)$.
This quantity can be regarded as the evaluation of a connection form on a vector field $\dot{w}$ restricted to the curve $w(t)$.
This connection is called the Uhlmann connection.

Now according to Eq.(35) of \cite{kinaiak} 
the pull-back of Uhlmann's connection by a local section provided by the polar decomposition $W=\varrho^{1/2}U$ is a one form which is given by the explicit formula\footnote{By an abuse of notation we denote the connection form on the bundle and its pull back to the base by the same letter $\mathcal A$.}
\begin{equation}
{\mathcal A}=-\sum_{I\neq J}\frac{(\sqrt{\lambda_I}-\sqrt{\lambda_J})^2}{
\lambda_I+\lambda_J}\vert I\rangle\langle I\vert d\vert J\rangle\langle J\vert, \qquad
d\equiv dx^{j}\partial_{j}
\label{explicitconnection}
\end{equation}
Moreover for the operator of parallel translation
we have
\begin{equation}
{\mathcal G}=\sum_{I,J}\frac{1}{
\lambda_I+\lambda_J}\vert I\rangle\langle I\vert d\varrho\vert J\rangle\langle J\vert, \qquad
d\equiv dx^{j}\partial_{j}
\label{explicitparallel}
\end{equation}
Here, it has been assumed that $\varrho=\sum_I\lambda_I\vert I\rangle\langle I\vert$ where generally both the eigenstates and the eigenvalues depend on some set of parameters $x^j$ parameterizing $\mathcal D$.

It is important to realize that in Eq.(\ref{explicitconnection}) ${\rm Tr}(\mathcal A)=0$. This means that $\mathcal A$ which is expected to be an $u(n)$-valued one form is rather an $su(n)$-valued one. It is known that Uhlmann's connection is reducible to the $SU(n)$ subbundle of our $U(n)$ bundle\cite{Rudolph}
which is obtained by restricting $W\in GL(n)$ to the subset of purifications also satisfying ${\rm Det W}\in {\mathbb R}_+$.

Let us illustrate this point when our density matrices are taken from our special set ${\mathcal B}\subset{\mathcal D}$ of Eq.(\ref{Ballset}). In this case $n=2^N$ and  $x:=u\in {\mathbb B}^{2N}$.
The (\ref{purify}) and (\ref{purify2}) purifications 
originally satisfy $W\in GL(2^N)$ with their determinants according to Eq.(\ref{asfollows}) being generally complex.
However, in the special case when $\tau=0$ or equivalently $\chi=0$, for the determinant we have ${\rm Det} W\in {\mathbb R}_+$ which is real and positive.
Then for sections of the $SU(2^N)$ subbundle we have the  polar decomposition in the form $W=\varrho^{1/2}S$ where $S\in SU(2^N)$.

Using the explicit expression (\ref{explicitconnection}) and (\ref{explicitparallel}) one can calculate the Uhlmann connection and parallel translation operators for our subclass $\mathcal B$.
The result is
\begin{equation}
  {\mathcal A}=\frac{1}{4}(1-{\rm sech}(\beta))[\slashed{n},d\slashed{n}]  
\label{Uconnection2}
\end{equation}
\begin{equation}
{\mathcal G}=\frac{1}{2}\left[\tanh\beta d\beta I+\slashed{n}d\beta-\tanh\beta d\slashed{n}\right]    
\label{Upara}
\end{equation}
The details of the calculation can be found in Appendix 15.3.

In the (\ref{density}) parametrization we have ${\bf u}=-\tanh\beta{\bf n}$, and the density matrix and the Uhlmann connection do not depend on the parameter $\chi$.
An alternative formula featuring the coordinates $\tau,r,{\bf n}$ is given by the expression
\begin{equation}
    {\mathcal A}=\frac{1}{4}\left(1-\sqrt{\frac{(1-r^2)^2+(1+\tau^2)^2-1}
{(1+r^2)^2+(1+\tau^2)^2-1}}\right)
[\slashed {n},
    d\slashed {n}]
\label{Uconnection3}
\end{equation}
If we take the restriction to the $\tau=0$ slice
\begin{equation}
    {\mathcal A}\vert_{\tau=0}=\frac{1}{2}\frac{r^2}{1+r^2}
[\slashed {n},
    d\slashed {n}]
\label{Uconnection4}
\end{equation}

In closing, we remark that an interesting formula can be obtained for the expectation value of the parallel translation operator. It is
\begin{equation}
\langle 2G\rangle_{\varrho}=dS-\beta d\langle\slashed n\rangle_{\varrho}    
\end{equation}
Here $dS$
is the change in the (\ref{freeenergy}) entanglement entropy under the change of our parameters.
Since $T=1/\beta$ and $\slashed{n}$ can be regarded as our Hamiltonian one can realize that $-2\langle G\rangle_{\varrho}$ equals to the combination $\frac{1}{T}dE-dS$.

\section{The physical meaning of the Uhlmann connection}

One can clarify the meaning of the Uhlmann connection by 
finding new coordinates for our $2N+2$ dimensional parameter space 
${\mathbb B}^{2N}\times S^1$ 
with  coordinates $(\chi, \beta, {\bf n})$. Alternately, one can use $(\tau, r, {\bf n})$ which gives the coordinates for the Euclidean plane ${\mathbb R}^{2N+2}$. The relationship between these coordinates is given by Eqs.(\ref{coord12}).
The new coordinates are Cartesian ones $(X_0,{\bf X},X_{2N+2})\in{\mathbb R}^{2N+3}$ constrained to lie on the surface of a $2N+2$ dimensional unit sphere i.e.
\begin{equation}
    X_0^2+\vert{\bf X}\vert^2+X_{2N+2}^2=1
\end{equation}
The corresponding coordinate transformation can be found in Section 15.4 of the Appendix. There one can see that the transformation relating 
this sphere $S^{2N+2}$ and ${\mathbb R}^{2N+2}$ is just stereographic projection from a pole of 
$S^{2N+2}$.

Define now the following family of $2^{N+1}\times 2^{N+1}$ unitary matrices parametrized by the points of $S^{2N+2}$. 
\begin{equation}
    U(X)=
    \frac{1}{\sqrt{2(1+X_{2N+2})}}
\begin{pmatrix}(1+X_{2N+2})I&-X_0I +i \slashed{\bf X}\\
    X_0 I+i\slashed{\bf X}&(1+X_{2N+2})I\end{pmatrix}
\label{first1}
\end{equation}
or alternatively using the (\ref{angles1})-(\ref{angles2}) transformation
\begin{equation}
    U(X)=    
    \frac{1}{\sqrt{1+\tau^2+r^2}}\begin{pmatrix}I&-\tau I+ir \slashed{\bf n}\\
    \tau I+ir\slashed{\bf n}&I\end{pmatrix}
\label{first2}
\end{equation}

Introducing the notation
\begin{equation}
Q_0:=\frac{1}{\sqrt{1+\tau^2+r^2}}I=\sqrt{\frac{1}{2}(1+X_{2N+2})}I
\end{equation}
\begin{equation}
Q_1:=\frac{1}{\sqrt{1+\tau^2+r^2}}(\tau I+ir \slashed{\bf n})=\frac{1}{\sqrt{2(1+X_{2N+2})}}(X_0I+i\slashed{\bf X})
\end{equation}
the compact form of $U$ is
\begin{equation}
U=\begin{pmatrix}Q_0&-Q_1^{\dagger}\\Q_1&Q_0\end{pmatrix}
\label{uuu}
\end{equation}
Now the (\ref{purify}) purification
of the density operator (\ref{alter}) can be written in the form
\begin{equation}
    W=\frac{1}{\sqrt{2^N}}(Q_0+iQ_1)
\label{pur1}
\end{equation}
Note that the columns of the matrix $U$
are comprising two sets of normalized $2^{N+1}$ component vectors. These two sets taken together form a basis in ${\mathbb R}^{2^{N+1}}$. Then the density operator has the form

\begin{equation}
    \varrho=WW^{\ast}=\frac{1}{2^N}(I+i(Q_1Q_0^{\ast}-Q_0Q_1^{\ast})=
    \frac{1}{2^N}\left(I-\frac{2r}{1+\tau^2+r^2}\slashed{\bf n}\right)
\end{equation}
Clearly since $2r=2\Omega\sinh\beta$ and
$1+\tau^2+r^2=2\Omega\cosh\beta$ we get back to (\ref{alter}).
Also comparing with the original (\ref{density}) parametrization we have
\begin{equation}
    {\bf u}=-\frac{2r}{1+\tau^2+r^2}{\bf n}=-\vert {\bf X}\vert {\bf n}=-\tanh\beta{\bf n}
\label{szokasos}
\end{equation}

Clearly, another purification can be defined using the other subspace spanned by the $2^N$ vectors hidden in $(-Q_1^{\dagger},Q_0)^T$.
In this case we have 
\begin{equation}
    W^{\prime}=\frac{1}{\sqrt{2^N}}(-Q_1^{\dagger}+iQ_0)=i
    \frac{1}{\sqrt{2^N}}(Q_0+iQ_1^{\dagger})
\label{pur2}
\end{equation}
hence up to the phase factor $i$ this change amounts to the reflection: ${\bf n}\mapsto -{\bf n}$.

Let us now define the anti-Hermitian part of a $2^N\times 2^N$ matrix
$M$ as ${\rm Im}(M):=\frac{1}{2}(M-M^{\dagger})$.
Now from Eq. (18) of \cite{LPBorn} one can see that with the definition $X:=X_0I-i\slashed{\bf X}$
the gauge-fields 
\begin{equation}
A_-=
\frac{{\rm Im}(XdX^{\dagger})}{2(1+X_{2N+2})}
\end{equation}
\begin{equation}
A_+=
\frac{{\rm Im}(X^{\dagger}dX)}{2(1+X_{2N+2})}
\label{Instantonquat}
\end{equation}
give rise to the higher dimensional monopoles familiar from \cite{Zalanek}
with monopole charge $\mp 2^{N-1}$.
Such objects are generalizing the well-known Dirac magnetic monopole\cite{Dirac} and in the $N=1$ case they corresponds to the 
instanton and anti instanton that are localized finite energy self-dual and anti self-dual solutions of the Euclidean $SU(2)$ Yang-Mills equations\cite{Belavin,Yang}.
Geometrically they are arising from the block diagonal elements of the Maurer-Cartan form 
$\omega:=U^{\dagger}dU$.

Now we would like to
find the relation between the monopole gauge fields $A_{\pm}$ and the Uhlmann gauge field ${\mathcal A}$ of Eq.(\ref{Uconnection4}).
What is the geometric relationship between these objects?
A calculation carried out in Appendix 2. shows that
\begin{equation}
 {\mathcal A}\vert_{\tau=0}=A_{\pm}\vert_{\tau=0}   
\label{UhlmannInstant}
\end{equation}
Hence on the $\tau=0$ surface Uhlmann's connection coincides with the higher dimensional monopole gauge-fields. Moreover, the corresponding $\tau$ independent part is not depending on the sign of the monopole charge. 

Notice also that with the choice
\begin{equation}
g:=\frac{\tau I-i\slashed{\bf r}}{\sqrt{\tau^2+r^2}} 
\in SU(2^N)
\end{equation}
the (\ref{Instantonquat}) gauge-fields can be written in the alternative form (higher dimensional instanton and anti-instanton gauge-fields)
\begin{equation}
A_+=\frac{\tau^2+r^2}{1+\tau^2+r^2}g^{\dagger}dg,\qquad
A_-=\frac{\tau^2+r^2}{1+\tau^2+r^2}gdg^{\dagger}
\end{equation}
In the limit $R\to \infty$ where $R:=\sqrt{\tau^2+r^2}$ they have the pure gauge form
\begin{equation}
\lim_{R\to\infty}A_+=g^{\dagger}dg,\qquad
\lim_{R\to\infty}A_-=gdg^{\dagger}
\end{equation}
One can check that the curvature two-forms $F_{\pm}=dA_{\pm}+A_{\pm}\wedge A_{\pm}$ are nonzero but 
\begin{equation}
    \lim_{R\to\infty}F_{\pm}=0
\end{equation}
Using these observations one can define
\begin{equation}
h:=g\vert_{\tau=0}=-i\slashed{\bf n}\in SU(2^N)    
\end{equation}
and using this we can write that
\begin{equation}
{\mathcal A}\vert_{\tau=0}=\frac{r^2}{1+r^2}h^{\dagger}dh    
\end{equation}

\section{Uhlmann anholonomy as a sequence of Thomas rotations}

From \cite{LPThomas} we know that for $N=1$ the $SU(2)$ part of Uhlmann's anholonomy is given by the product of a sequence of $SU(2)$ matrices forming a representation of Thomas rotations familiar from special relativity. 
Now we would like to see how this interpretation survives also in the general $N>1$ case.

The success of proving this for $N=1$ rests in the very special nature of this
case. For $N=1$ we have $2\times 2$ matrices hence the direct calculation of matrix square roots needed in equations like Eq.(\ref{yij})  is straightforward.
This cannot be expected for $N>1$ case.
However, one can also notice the peculiarity that in the  $N=1$ case the square roots of density matrices gives the usual $SL(2,{\mathbb C})$
spinor representations of Lorentz boosts familiar from four dimensional Minkowski space-time geometry.
As is well-known the product of two boosts gives rise to a third boost times an $SU(2)$ rotation. Hence the parameters of this new boost and rotation can explicitly be written down\cite{Mukunda} and then related to Uhlmann's anholonomy\cite{LPThomas}.

Now due to the special nature of our (\ref{density}) density matrices it is natural to suspect that this interpretation of Uhlmann's anholonomy will be true even for arbitrary $N$.
The reason for this is that the square roots of the density matrices still have the interpretation of Lorentz boosts in higher dimension, and Thomas rotation is still arising from the product of two boosts. Then one can hope that it is possible to identify this rotation (the Wigner rotation) as Uhlmann's anholonomy. In order to explicitly see that this can be done we proceed as follows.

First we note that the basic building block of Uhlmann's anholonomy is the $2^N\times 2^N$ unitary matrix 
\begin{equation}
U_{ij}:={\varrho_i}^{-1/2}{\varrho_j^{-1/2}}\left(\varrho_j^{1/2}\varrho_i
\varrho_j^{1/2}
\right)^{1/2}
\label{final}
\end{equation}
familiar from Eq.(\ref{y}).
In order to be able to manipulate expressions like the one in Eq.(\ref{final}) it is useful to rather consider $2^{N+1}\times 2^{N+1}$ matrices of the form
\begin{equation} L(u):=\cosh\frac{\beta_u}{2} 1+\sinh\frac{\beta_u}{2}\hat{\bf u}\boldsymbol{\gamma}\gamma^0
\label{back}
\end{equation}
where $\hat{\bf u}$ is a unit vector and the symbol $1$ refers to the $2^{N+1}\times 2^{N+1}$ unit matrix.

 The new gamma matrices showing up here are the
 $2^{N+1}\times 2^{N+1}$ dimensional
 ones living in $2N+3$ dimension. They are generated from the familiar $2^N\times 2^N$ gamma matrices in $2N+1$ dimension via the recursive procedure of Appendix 15.1. The novelty here that now we have 
\begin{equation}
\{\gamma^{\mu},\gamma^{\nu}\}=2\eta^{\mu\nu}1,\qquad \mu,\nu=0,1,2,\dots,2N+2
\end{equation}
i.e. we have switched from the Euclidean signature to the Lorentzian one.
In Eq.(\ref{back}) we used from this Lorentzian  set merely the first $2N+2$ ones, namely $\gamma^0$ and $\gamma^k$ with $k=1,2,\dots ,2N+1$.
In terms of our familiar $2N+1$ dimensional gamma matrices one has the explicit form
\begin{equation}
\gamma^0=\sigma_1\otimes iI,\qquad \gamma^k=\sigma_2\otimes {\Gamma}_k    
\label{kisgamma}
\end{equation}

The upshot of these considerations is that in this way we can  facilitate a Lorentz boost interpretation for $L(u)$ showing up in Eq.(\ref{back}). 
Indeed, $\gamma^k\gamma^0=\sigma_3\otimes\Gamma_k$
hence
\begin{equation}
L(u):=\cosh\frac{\beta_u}{2} {\bf 1}\otimes I+\sinh\frac{\beta_u}{2}\sigma_3\otimes {\hat{\bf u}}\boldsymbol{\Gamma}
\label{back2}
\end{equation}
where ${\bf 1}$ is the $2\times 2$ unit matrix, see Section 15.1. in the Appendix. One can then realize that $L(u)$ has a block diagonal structure, containing two $2^N\times 2^N$ blocks. Moreover, the matrix in the left hand side of Eq.(\ref{back}) has the form

\begin{equation}
{L(u)}
=\begin{pmatrix}Z_u^{1/2}\varrho_u^{1/2}&0\\0&Z_u^{-1/2}\varrho_u^{-1/2}\end{pmatrix}
\label{back3}
\end{equation}
where $Z_u$ is the partition function familiar from Eqs.(\ref{zefi})-(\ref{zefi2}).
(See also Eq. (\ref{para}) of Appendix 15.2.) 
Hence $L(u)$ contains the canonical purification of our density matrix and its inverse.

The meaning of the two blocks in (\ref{back3}) is also easy to understand.
According to Eqs.(\ref{pur1}) and(\ref{pur2}) we have the possibility of forming two different purifications originating from the different columns of the $2^{N+1}\times 2^{N+1}$ unitary matrix of Eq.(\ref{uuu}). They differ only in the sign of the vector  ${\bf u}$.
This sign change boils down to either considering $\varrho_u$ or its inverse.

In Section 15.5. of the Appendix we show that the matrix defined as
\begin{equation}
    {\mathcal U}_{uv}:=L^{-1}(u)L^{-1}(v)(L(v)L(u)L(u)L(v))^{1/2}=\begin{pmatrix}
    U_{uv}&0\\0&U_{uv}    
    \end{pmatrix}
\label{idcopy}
\end{equation}
contains two identical copies
of our basic building block.
In particular we have the explicit formula 
\begin{equation}
{\mathcal U}_{uv}=\cos\left(\frac{\delta}{2}\right)1+
\sin\left(\frac{\delta}{2}\right)
\frac{2\hat{u}_j\hat{v}_k\Sigma^{jk}}{\sqrt{1-(\hat{\bf u}\hat{\bf v})^2}},\qquad\Sigma^{jk}:=\frac{1}{4}[\gamma^j,\gamma^k]
\label{mesterke}
\end{equation}
where
\begin{equation}
\cos\left(\frac{\delta}{2}\right)=
\frac{\cosh\left(\frac{\beta_u}{2}\right)
\cosh\left(\frac{\beta_v}{2}\right)+
\sinh\left(\frac{\beta_u}{2}\right)
\sinh\left(\frac{\beta_v}{2}\right)(\hat{\bf u}\hat{\bf v})
}{\cosh\left(\frac{\beta_w}{2}\right)}
\label{lenyeg}
\end{equation}
\begin{equation}
\sin\left(\frac{\delta}{2}\right)=
{\sqrt{1-(\hat{\bf u}\hat{\bf v})^2}}
\frac{
\sinh\left(\frac{\beta_u}{2}\right)
\sinh\left(\frac{\beta_v}{2}\right)
}{\cosh\left(\frac{\beta_w}{2}\right)}
\end{equation}
Here it is understood that 
$\cosh\left(\frac{\beta_w}{2}\right)$ have to be expressed in terms of 
$\beta_u$ and $\beta_v$ from 
\begin{equation}    \cosh\beta_w=\cosh\beta_u\cosh\beta_v+\sinh{\beta}_u\sinh\beta_v(\hat{\bf u}\hat{\bf v})   
\label{kifejez2}
\end{equation}
Notice also that due to the (\ref{kisgamma}) special structure of the $2^{N+1}\times 2^{N+1}$  $\gamma^k$ matrix we have
$\Sigma^{jk}=\frac{1}{4}I\otimes [\Gamma_j,\Gamma_k]$ hence indeed we are having two identical copies of $U_{uv}$ inside ${\mathcal U}_{uv}$. They are featuring our original $2^N\times 2^N$ gamma matrices. Hence the explicit formula for $U_{uv}$ is obtained from 
(\ref{mesterke}) by replacing ${\gamma^k}$ by
$\Gamma_k$.

The reason for having two identical copies inside (\ref{idcopy}) rests in Eq.(\ref{UhlmannInstant}).
This equation says that no matter whether we use the instanton or the anti-instanton, the Uhlmann connection can be derived from both of them by taking the $\tau\to 0$ limit.
Alternatively one can say that the sign of the topological charge of the corresponding higher dimensional monopoles does not makes its presence in the Uhlmann connection.
In the language of density matrices this means that 
the sign of ${\bf n}$ which makes the difference between $\varrho_u$ and $\varrho_u^{-1}$ is not important when considering Uhlmann's anholonomy.

Let us now consider the quantity
\begin{equation}
    iB:=\frac{2\hat{u}^j\hat{v}^k\Sigma_{jk}}{\sqrt{1-(\hat{\bf u}\hat{\bf v})^2}}
\label{ide1}
\end{equation}
Then one can check that $B^{\ast}=B$.
One can also see that we also have the property $B^2=I$ then ${\mathcal U}_{uv}$ can be written in the form
\begin{equation}
    {\mathcal U}_{uv}=e^{i(\delta/2)B}
\label{ide2}
\end{equation}
Now an important comment is in order. The $\Sigma_{jk}$ operators showing up in (\ref{mesterke}) are generators of the group $Spin(2N+1)$ the two-fold cover of the rotation group $SO(2N+1)$. 
Then the matrices $U_{uv}$ for any pair (${\bf u}$,${\bf v}$) are forming the $2^N$ dimensional spinor representation of this rotation group. Since for $N>1$ we have  $Spin(2N+1)\subset SU(2^N)$
we are not expecting to be able to obtain arbitrary $SU(n)$ quantum gates with $n=2^N$ via Uhlmann anholonomy.
This is as it should be since our density matrices are very special.

Notice that in the special case $N=1$
\begin{equation}
B=\frac{\hat{\bf u}\times\hat{\bf v}}{\vert\hat{\bf u}\times\hat{\bf v}\vert }\boldsymbol{\sigma}
\end{equation}
hence with the definition
\begin{equation}
    {\bf n}:=\frac{\hat{\bf u}\times\hat{\bf v}}{\vert\hat{\bf u}\times\hat{\bf v}\vert }
\end{equation}
we have
\begin{equation}
    U_{uv}=e^{i(\delta/2){\bf n}\boldsymbol{\sigma}}
\label{ipsilon}
\end{equation}
which is the canonical form of the $SU(2)$ representation of a rotation
with rotation angle $\delta$ and axis of rotation ${\bf n}$.
In this case we have the coincidence $Spin(3)\simeq SU(2)$ hence all $SU(2)$ gates can be generated by Uhlmann anholonomy. But this is not a surprise, since in this case we can have a Bloch ball parametrization for {\it all} nondegenerate density matrices inside the ball.

Another interesting possibility is arising in the $N=3$ case.  Here one can consider the $Spin(6)\subset Spin(7)\subset SU(8)$ chain.
However, it is known that 
$Spin(6)\simeq SU(4)$ hence using Uhlmann's anholonomy there is the possibility of describing special two-qubit gates inside three qubit ones. This will be discussed in more detail later.
Finally we notice that in the general case with $N\geq 1$ the angle of rotation can also be expressed as
\begin{equation}
    \tan\left(\frac{\delta}{2}\right)=
    \frac{\sqrt{\vert\bf u\vert^2
    \vert\bf v\vert^2-({\bf u}{\bf v})^2}}{(1+C_u)(1+C_v)+{\bf u}{\bf v}},\qquad {\bf u}=\tanh\beta_u\hat{\bf u},\qquad C_u:=\frac{1}{\cosh\beta_u}
\label{meghat}
\end{equation}
One can then even further simplify this by introducing the quantities familiar from Eq. (19)
of Ref.\cite{LPThomas}
\begin{equation}
    {\bf a}:=\frac{\bf u}{1+C_u}=\tanh(\beta_u/2)\hat{\bf u},\qquad
    {\bf b}:=\frac{\bf v}{1+C_v}=\tanh(\beta_v/2)\hat{\bf v}
\label{atteres}
\end{equation}
\begin{equation}
    \tan\left(\frac{\delta}{2}\right)=
    \frac{\sqrt{a^2b^2-({\bf a}{\bf b})^2}}{1+{\bf a}{\bf b}}
\end{equation}
As shown in the Appendix these quantities are related to the stereographically projected coordinates of Eq.(\ref{sproj}) as ${\bf X}_u=-{\bf u}$ and ${\bf r}_a=-{\bf a}$.

Then an alternative expression for $\cos{\delta/2}$ can be given in the form
\begin{equation}
\cos{\delta/2}=\frac{1+{{\bf a}{\bf b}}}{\sqrt{1+2{\bf ab}+a^2b^2}}
\end{equation}
And now the (\ref{mesterke}) Thomas rotation
can also be written in the form
\begin{equation}
U_{uv}:=U_{ab}=\frac{I+\slashed{a}\slashed{b}}
{\sqrt{1+2{\bf ab}+a^2b^2}}=
\frac{(1+{\bf ab})I+\frac{1}{2}[\slashed{a},\slashed{b}]}
{\sqrt{1+2{\bf ab}+a^2b^2}}
\label{hyppar}
\end{equation}
This (\ref{atteres}) "half angle" parametrization related to stereographic projection will turn out to be convenient in our next section.

\section{Uhlmann anholonomy for geodesic triangles}

Now we consider anholonomy effects for density matrices of the (\ref{density}) type.
As we have seen these $N$ qubit density matrices are parametrized by the points of  
the interior of the unit ball ${\mathbb B}^{2N}$. 
Now an application of the distance formula of (\ref{Buresertem})
to two infinitesimally separated density matrices of full rank  $\varrho$ and $\varrho+d\varrho$ results in a line element of the Bures metric\cite{Bures}. The pull back of the Bures metric living on the space ${\mathcal D}$ of nondegenerate density matrices also defines a metric on ${\mathbb B}^{2N}$.
A calculation of this metric is carried out in Section 15.6. of the Appendix. This shows that the corresponding line element  can be written in the following form
\begin{equation}
    ds^2=g_{jk}du^jdu^k=\frac{1}{4\cosh^2\beta_{u}}(d\beta_{{u}}^2+\sinh^2\beta_u (d\mathbf{\Hat{u}})^2)=\frac{1}{4\cosh^2\beta_{u}} ds^2_{\mathcal{H}^{2N+1}}
\label{hyphyp}
\end{equation}
An alternative expression for this line element can be written down using the parallel translation operator of Eq.(\ref{Upara})
\begin{equation}
ds^2=\langle {\mathcal G}^2\rangle_{\varrho}={\rm Tr}({\mathcal G}^2\varrho)    
\end{equation}
In these expressions $ds^2_{\mathcal{H}^{2N+1}}$ is the metric on the upper sheet of the $2N+1$ dimensional double sheeted hyperboloid. Hence the Bures metric is conform equivalent to this hyperbolic metric.
Note that $(d\mathbf{\Hat{u}})^2)$ denotes the metric
on the $2N$ dimensional sphere $S^{2N}$.

Now it is sensible to consider geodesic triangles in ${\mathbb B}^{2N}$
with respect to this metric. Such triangles are defined by three points ${\bf u}, {\bf v}, {\bf w}\in{\mathbb B}^{2N}$. 
In the following we will be interested in deriving an explicit formula for the Uhlmann anholonomy around such a geodesic triangle.

If we define three not necessarily orthogonal unit vectors ${\bf e}_u, {\bf e}_v, {\bf e}_w$ then one can write
\begin{equation}
{\bf u}=-\tanh\beta_u {\bf e}_u,\qquad
{\bf v}=-\tanh\beta_v {\bf e}_v,\qquad
{\bf w}=-\tanh\beta_w {\bf e}_w
\label{elsoverzio}
\end{equation}.
Clearly all three density matrices coming from the
parametrization above can be given the thermal state form of (\ref{alter}).
For example
\begin{equation}
    \varrho_u=\frac{1}{2^N}[I-\tanh\beta_u H_u]
=\frac{{e^{-\beta_u H_u}}}{{\rm Tr}(e^{-\beta_u H_u})}
,\qquad H_u:={\boldsymbol{\Gamma}} {\bf e_u}=\slashed{e}_u
\end{equation}

It is rewarding to introduce the rescaled vectors
of the previous section
\begin{equation}
  {\bf a}=-\tanh\frac{\beta_u}{2}{\bf e}_u,\qquad
{\bf b}=-\tanh\frac{\beta_v}{2}{\bf e}_v,\qquad
{\bf c}=-\tanh\frac{\beta_w}{2}{\bf e}_w
\label{renormalized}
\end{equation}
In terms of these new vectors
for $N=1$ the Uhlmann anholonomy for geodesic triangles has been calculated in \cite{LPThomas}.
Based on the results of Section 7. It is clear that this result of Ref.\cite{LPThomas}
can be generalized for the (\ref{Ballset}) set of $N$-qubit density matrices as well.
In the notation of Eqs. (\ref{nice}) and (\ref{final})
one gets 
\begin{equation}
    {\mathcal R}({\bf a},{\bf b},{\bf c}):=U_{13}U_{32}U_{21}=U_{uw}U_{wv}U_{vu}=U_{ac}U_{cb}U_{ba}
\label{rewritehyp}
\end{equation}
Here it is understood that we have to use the (\ref{atteres}) definitions with $\hat{\bf u}={\bf e}_u$ etc. and for the explicit form of ${\mathcal R}({\bf a},{\bf b},{\bf c})$ the result of Eq.(29) of Ref.\cite{LPThomas} generalizes as follows 
\begin{equation}
    {\mathcal R}({\bf a},{\bf b},{\bf c})=\frac{(1+{\bf pq})I+\frac{1}{2}[\slashed{p},\slashed{q}]}{\sqrt{1+2{\bf pq}+p^2q^2}},\qquad 
    \slashed{p}:={\bf p}{\boldsymbol\Gamma},\qquad
    \slashed{q}:={\bf q}{\boldsymbol\Gamma}
\label{mester}
\end{equation}
where
\begin{equation}
\frac{\bf p}{p^2}={\bf a}+\frac{1+a^2}{\vert{\bf c}-{\bf a}\vert^2}({\bf c}-{\bf a}),\qquad
\frac{\bf q}{q^2}={\bf a}+\frac{1+a^2}{\vert{\bf b}-{\bf a}\vert^2}({\bf b}-{\bf a})
\label{inversion1}
\end{equation}
\begin{equation}
p^2:=\vert{\bf p\vert^2}=\frac
{\vert{\bf c}-{\bf a}\vert ^2}{1+2{\bf ac}+a^2c^2},\qquad
q^2:=\vert{\bf q\vert^2}=\frac{\vert{\bf b}-{\bf a}\vert ^2}{1+2{\bf ab}+a^2b^2}
\label{inversion2}
\end{equation}
The meaning of the (\ref{inversion1}) transformations is easy to understand. They are combinations of two inversions with respect to spheres. For example the right hand side of the first of (\ref{inversion1}) is an inversion with respect to a sphere with radius $1+a^2$ centered at ${\bf a}\in {\mathbb R}^{2N+1}$. Its left hand side is an inversion with respect to a spere of radius $1$, centered at the origin.
Notice the special role played by the vector {\bf a} in both expressions. It is serving as the starting point for the traversal of the geodesic triangle.

Let us expand the vectors ${\bf p}$ and ${\bf q}$ with respect to the normalized (but not necessarily orthogonal) basis vectors $({\bf e}_u,{\bf e}_v,{\bf e}_w):= ({\bf e}_1,{\bf e}_2,{\bf e}_3)$
\begin{equation}
{\bf p}=p^a{\bf e}_a,\qquad {\bf q}=q^b{\bf e}_b,\quad a,b=1,2,3
\end{equation}
Then (the raising and lowering of indices is effected by $\delta^{ab}$ and $\delta_{ab}$) for the commutator term of (\ref{mester}) we have
\begin{equation}
\frac{1}{2}[\slashed{p},\slashed{q}]=\frac{1}{2}(p^aq^b-p^bq^a){\slashed{\bf e}_a}{\slashed{\bf e}_b}=\frac{1}{2}\varepsilon^{abc}({\bf p}\times{\bf q})_c{\slashed{\bf e}_a}{\slashed{\bf e}_b}=i({\bf p}\times{\bf q}){\boldsymbol{\Sigma}}
\label{kommutator12}
\end{equation}
where
\begin{equation}
    ({\bf p}\times{\bf q})_c=\varepsilon_{abc}p^aq^b,\qquad
    i{{\Sigma}}^c=\frac{1}{2}\varepsilon^{abc}
    \slashed{\bf e}_a\slashed{\bf e}_b=\frac{1}{4}
\varepsilon^{abc}e^{j}_ae^{k}_b[\Gamma_{j},\Gamma_{k}]:=\varepsilon^{abc}e^{j}_ae^{k}_b\Sigma_{jk}
\end{equation}
Here $\Sigma_{jk}=\frac{1}{4}[\Gamma_j,\Gamma_k]$ where $j,k=1,2,\dots,2N+1$
generate a $spin(2N+1)$ algebra, where $Spin(2N+1)$ is the double cover of the rotation group $SO(2N+1)$. It is well-known that $Spin(3)\simeq SU(2)$ and $Spin(6)\simeq SU(4)$.
In terms of the generators $\Sigma_{jk}$ the commutation relations of the $spin(2N+1)$ algebra take the following form 
\begin{equation}
[\Sigma_{jk},\Sigma_{mn}]=\delta_{jn}\Sigma_{km}
+\delta_{km}\Sigma_{jn}-\delta_{jm}\Sigma_{kn}
-\delta_{kn}\Sigma_{jm}
\label{algebra}
\end{equation}

In order to better understand expression (\ref{mester}) we recall that
the Bures distance between two density matrices
$\varrho_u$ and $\varrho_v$
can be expressed using the fidelity $F({\bf u},{\bf v})$. Namely according to (\ref{Wasserstein}) this distance squared is given by
\begin{equation}
    d_B^2(\varrho_u,\varrho_v)=2\left(1-{\rm Tr}(\varrho_u^{1/2}\varrho_v\varrho_u^{1/2})^{1/2}\right)=
    2\left(1-\sqrt{F({\bf u},{\bf v})}\right)
    \label{buresdistance}
\end{equation}
Now we claim that the fidelity is
\begin{equation}
 F({\bf u},{\bf v})=\frac{1}{2}\left(1+{\mathcal C}_u{\mathcal C}_v+{\bf uv}\right)=1-\frac{\vert {\bf a}-{\bf b}\vert^2}{(1+a^2)(1+b^2)}
\label{fidelity}
\end{equation}
where for the definition of $\mathcal C_u$, ${\mathcal C}_v$ see Eq. (\ref{meghat}).
In order to see this just recall Eqs. (\ref{back}), (\ref{kifejez}),
(\ref{elw}) and (\ref{square}).
Then we arrive at
\begin{equation}
 \rm{Tr}(\varrho_u^{1/2}\varrho_v\varrho_u^{1/2})^{1/2}=\rm{Tr}\left({\frac{L(w)}{2^{N}
 \sqrt{\cosh\beta_u\cosh\beta_v}}}\right)=
 \frac{\cosh\beta_w/2}
{\sqrt{\cosh\beta_u\cosh\beta_v}}=
\sqrt{\frac{1+\cosh\beta_w}{2\cosh\beta_u\cosh\beta_v}}
\nonumber
\end{equation}
Now using Eq.(\ref{kifejez}) one obtains (\ref{fidelity}).

A straightforward calculation using (\ref{kommutator12}) and (\ref{fidelity}) then shows that
\begin{equation}
     {\mathcal R}({\bf a},{\bf b},{\bf c})=\cos\left(\frac{\delta}{2}\right)I+i\sin\left(\frac{\delta}{2}\right)({\bf m}{\boldsymbol{\Sigma}})\in Spin(2N+1)
\label{forgatasos}
\end{equation}
where
\begin{equation}
    \cos\left(\frac{\delta}{2}\right)=\frac{F({\bf u},{\bf v})
    +F({\bf v},{\bf w})+F({\bf w},{\bf u})-1}
    {2\sqrt{F({\bf u},{\bf v})F({\bf v},{\bf w})F({\bf w},{\bf u})}},\qquad
    {\bf m}:=\frac{{\bf p}\times{\bf q}}{\vert{\bf p}\times{\bf q}\vert}
\label{crossszorzatos}
\end{equation}
Where here it is understood that the triple
$({\bf u},{\bf v},{\bf w})$
should be expressed in terms of the one $({\bf a},{\bf b},{\bf c})$, see also Eqs. (\ref{inversion1}) and ({\ref{fidelity}).}
From this we learn that the holonomy is a Thomas rotation belonging to the $Spin(2N+1)$ subgroup. 
This is the double cover of the $2n+1$ dimensional rotation group sitting inside the proper orthochronous Lorentz group $SO(2N+1,1)$.
The axis of the rotation is given by ${\bf m}$ and the angle of rotation is $\delta$.

Note however, that here we abused the notation since our vector ${\bf m}$ showing up in (\ref{forgatasos}) has three components i.e. ${\bf m}\leftrightarrow m^a$
and the vectors ${\bf p}$ and ${\bf q}$ are regarded in (\ref{crossszorzatos}) as three component vectors as well. But more precisely one should rather reinterpret them as ones like
$m^j=m^ae_a^j$ in ${\mathbb R}^{2N+1}$.
The (\ref{forgatasos}) way is emphasizing the fact that this $SO(2N+1)$ rotation is effected in a three dimensional subspace spanned by ${\bf e}_a$, $a=1,2,3$.
Indeed, one can check using Eq.(\ref{algebra}) and the Jacobi identity that in the special case when the ${\bf e}_a$ span an orthonormal basis, i.e. $e_a^je_b^k\delta_{jk}=\delta_{ab}$ we have
\begin{equation}
[\Sigma^a,\Sigma^b]=2i\varepsilon^{abc}\Sigma_c 
\end{equation}
i.e. $\Sigma^a$ are spanning an $su(2)\simeq spin(3)$ subalgebra. In this case the (\ref{mester}) rotation will be an ordinary three dimensional rotation matrix in the spinor representation.

\section{The physics of Uhlmann's anholonomy}

Now we would like to get some insight into the physical meaning of Uhlmann's anholonomy based on our thermo-field double (TFD) approach.
Our starting point is our basic result of Eq.(\ref{essence}) which for a geodesic triangle we rewrite as
\begin{equation}
(L_{ac}L_{cb}L_{ba}\otimes I)\vert\varphi_a\rangle =
(I\otimes R_{ab}R_{bc}R_{ca})\vert\varphi_a\rangle,\qquad
R_{ij}=\overline{U}_{ij}
\label{atiras}
\end{equation}
with $U_{ij}$ in hyperbolic parametrization having the form of Eq.(\ref{hyppar}).
Recall now that we have the (\ref{correspondence}) correspondence $\varrho_a^{1/2}\leftrightarrow \vert\varphi_a\rangle$.
However, we would like to implement a more general form of this by the polar decomposition of (\ref{polar1}) and (\ref{sjöquist})
\begin{equation}
W_a=\varrho_a^{1/2}U_a\leftrightarrow \vert \psi_a\rangle:=(I\otimes U_a^T)\vert\varphi_a\rangle=(I\otimes \overline{U_a}^{-1})\vert\varphi_a\rangle
\label{good}
\end{equation}
or alternatively (see Section 15.2 of the Appendix)
\begin{equation}
W_a=U_a\varrho_a^{1/2}\leftrightarrow \vert \psi_a\rangle:=(U_a\otimes I)\vert\varphi_a\rangle
\label{good2}
\end{equation}
hence we can also implement this unitary as a manipulation performed on the left subsystem.

The benefit of this generalization is that it also implements our extra parameter $\chi$ (or $\tau$) which shows up in $U_a$ in the form (see Eq.(\ref{sjöquist}))
\begin{equation}
U_a=e^{i\chi/2}\sqrt{\frac{\cosh\beta +1}{\cosh\beta +\cos\chi}}\left[\cos\left(\frac{\chi}{2}\right) I +i\sin\left(\frac{\chi}{2}\right)\slashed{a}\right]    
\label{ua}
\end{equation}
Moreover, we already know from our considerations with instantons of Section 7. that $\chi$ is playing the role of a compactified Euclidean time coordinate. The corresponding circle is showing up in the configuration space of our parameters in the form ${\mathbb B}^{2N}\times S^1$. In Section 15.4. of the Appendix we will show that this space is conformally equivalent to $S^{2N+2}$. It would be nice to see its physical role in our approach.

This implementation of $\chi$ can be taken into account by replacing $\vert\varphi_a\rangle$ with $\vert \psi_a\rangle$ in the left hand side of (\ref{atiras}). 
Recall now that
by virtue of Eqs. (\ref{geomean1}), (\ref{rewritehyp}), (\ref{mester}) and (\ref{forgatasos})
the form of $R_{ab}R_{bc}R_{ca}$ used on the right
can be written as
\begin{equation}
R_{ab}R_{bc}R_{ca}=\bar{U}_{ab}\bar{U}_{bc}\bar{U}_{ca}
\label{riri}
\end{equation}
It is easy to show that the implementation of $\chi$ modifies this to
\begin{equation}
\bar{U}_a^{-1}\bar{U}_{ab}\bar{U}_{bc}\bar{U}_{ca}
\label{riri2}
\end{equation}
Then the meaning of this modification in the right system is clear. This is a sequence of unitary manipulations familiar from quantum circuits of quantum computation. The last three of them are of the form of generalized Thomas rotations.
The first one is containing $\chi$.

What about the manipulations performed on the left hand side?
According to (\ref{geomean}) these are built up from positive operations featuring the geometric mean in the form $L_{ij}:=\varrho_i\sharp\varrho_j^{-1}$ with their physical meaning not at all clear. In the following we will argue that these operations can be interpreted as a sequence of filtering measurements performed step by step on our sequence of entangled states.

In order to show this
we employ yet another representation for our entangled states $\vert\psi_a\rangle$,
$\vert\psi_b\rangle$ and $\vert\psi_c\rangle$ playing the main roles in our considerations.
This representation is based on "block-spinors". These are the quantities featuring the columns of the unitary familiar from Eq.(\ref{uuu}). 
A block-spinor is simply an $2n\times n$ matrix with  $n=2^N$ of the form $(Q_0,Q_1)^T$ where $Q_0$ and $Q_1$ are complex $n\times n$ matrices. Geometrically such an object is defining an $n$-plane in ${\mathbb C}^{2n}$ spanned by the $n$ columns of $(Q_0,Q_1)^T$.
Then the correspondence between block-spinors, purifications and entangled states is as follows
\begin{equation}
\begin{pmatrix}Q_0\\Q_1\end{pmatrix}\leftrightarrow W=\frac{1}{\sqrt{2^N}}\left(Q_0+iQ_1\right)\leftrightarrow \vert\psi\rangle
\end{equation}
Since ${\varrho_a}^{1/2}=(I+\slashed{a})/\sqrt{{2^N}(1+a^2)}$ in our setup we explicitly have
\begin{equation}
    \vert a\rangle :=\frac{1}{\sqrt{1+a^2}}\begin{pmatrix}I\\-i\slashed{a}\end{pmatrix}U_a\longleftrightarrow W_a=\varrho^{1/2}_aU_a\longleftrightarrow \vert\psi_a\rangle
    \label{blockspin}
\end{equation}
Clearly the basis of this correspondence is the fact that
an $n$-plane in ${\mathbb C}^{2n}$ spanned by $n$ orthonormal basis vectors as columns of the block-spinor is defined up to right multiplication with respect to an $U(n)$ matrix. This gauge degree of freedom captures the freedom arising in the purification $W\mapsto WU$.

Now we introduce a inner product which is taking values in ${\mathbb M}(n)$ as
\begin{equation}
\langle P\vert Q\rangle:=P_0^{\ast}Q_0+P_1^{\ast}Q_1
\end{equation}
Then we call a block-spinor normalized when we have
$\langle Q\vert Q\rangle = I$.    

Now let us observe that
\begin{equation}
    U_b\langle b\vert a\rangle U_a^{\ast}=\frac{1}{\sqrt{1+b^2}\sqrt{1+a^2}}\begin{pmatrix}I,&i\slashed {b}\end{pmatrix}\begin{pmatrix} I\\-i\slashed{a}\end{pmatrix}=
    \frac{I+\slashed{b}\slashed{a}}{\sqrt{1+b^2}\sqrt{1+a^2}}
\end{equation}

Moreover, we also have
\begin{equation}
\vert\langle b\vert a\rangle\vert=\sqrt{\frac{1+2{\bf ab}+a^2b^2}{(1+a^2)(1+b^2)}} 
\end{equation}
hence finally using Eq.(\ref{hyppar})
\begin{equation}
    \frac{\langle b\vert a\rangle}{\vert\langle b\vert a\rangle\vert}=U_b^{\ast}\frac{I+\slashed{b}\slashed{a}}{\sqrt{1+2{\bf ab}+b^2a^2}}U_a=U_b^{\ast}U_{ba}U_a
\end{equation}
For geodesic triangles one can iterate this procedure
\begin{equation}
    \vert a\rangle \frac{\langle a\vert c\rangle}{\vert\langle a\vert c\rangle\vert}
\frac{\langle c\vert b\rangle}{\vert\langle c\vert b\rangle\vert}\frac{\langle b\vert a\rangle}{\vert\langle b\vert a\rangle\vert}=\varrho_a^{1/2}U_{ac}U_{cb}U_{ba}U_a
\label{crucial}
\end{equation}
Now we have (\ref{inverseunitary}) at our disposal hence
$U_{ac}U_{cb}U_{ba}=(U_{ab}U_{bc}U_{ca})^{-1}=(\overline{U_{ab}U_{bc}U_{ca}})^T$.
Finally the right hand side of (\ref{crucial}) in entangled state notation takes the form
\begin{equation}
(I\otimes \bar{U}_a^{-1}\overline{U_{ab}U_{bc}U_{ca}})\vert\varphi_a\rangle
\label{next}
\end{equation}

Let us now reinterpret the left hand side of Eq.(\ref{crucial}) in block-spinor language.
\begin{equation}
    \frac{\vert a\rangle\langle c\vert}{\vert\langle a\vert c\rangle\vert}
\frac{\vert c\rangle\langle b\vert}{\vert\langle c\vert b\rangle\vert}
\frac{\vert b\rangle\langle a\vert}{\vert\langle b\vert a\rangle\vert}\vert a\rangle=L_{ac}L_{cb}L_{ba}\varrho^{1/2}_aU_a 
=L_{ac}L_{cb}L_{ba}U_a\varrho^{1/2}_a
\label{crucial2}
\end{equation}
The block-spinor language clearly shows that one can interpret the left hand side of (\ref{crucial2}) as a sequence of filtering measurements acting on the initial state $\vert a\rangle$ interpreted now as an $n=2^{N}$ dimensional subspace in ${\mathbb C}^{2n}$. The filtering measurement is effected by applying the rank $n$ orthogonal projectors of the form $\vert b\rangle\langle a\vert$ and its successors
divided by the corresponding "transition probabilities" at each step.

At the level of the right hand side of Eq.(\ref{crucial2}) it is clear that these rank $n$ projectors representing filtering measurements also correspond to Hermitian positive operators of the form $L_{ba}=\varrho_b\sharp\varrho_a^{-1}\in{\mathbb P}(n)$.
The physical meaning of these operators have also been elucidated in the literature. These operators implement optimal measurements for distinguishing between arbitrary two given full rank states\cite{Fuchs,Ericsson} in this case $\varrho_a$ and $\varrho_b$.

More precisely, in our mixed state context one can ask for the most advantageous measurement for distinguishing the states in a statistical sense. This means that we need to make distinction between two associated probability distributions $p_i^{(a)}={\rm Tr }(E_i\varrho_a)$ and
$p_i^{(b)}={\rm Tr }(E_i\varrho_b)$.  Here the $E_i$ refers to a set of POVMs (positive operator valued measures). These are forming a complete set of nonnegative Hermitian operators acting on our left subsystem where $\sum_i E_i=I$.
One can then define the distance 
\begin{equation}
    D(\varrho_a,\varrho_b):=\max_{\{E_i\}}{\rm arccos}\left(\sum_i\sqrt{{\rm Tr} E_i\varrho_a}
    \sqrt{{\rm Tr} E_i\varrho_b}\right)
\label{maximum}
\end{equation}
which is based on the statistical distance of Bhattacaryya and Wootters\cite{Batta,Wootters}.
Now in Ref.\cite{Fuchs} it has been shown that the best measurement providing the maximum in (\ref{maximum})
is achieved by the observable 
    $M:=L_{ba}$. Moreover, in this case the maximum value is attained via the equation $\cos D(\varrho_a,\varrho_b)=\sqrt{F(a,b)}$ where $F(a,b)$ is the fidelity.
In addition in Ref.\cite{Ericsson} is was noticed that this maximum value coincides with the length of the geodesic segment (with respect to the Bures-Uhlmann metric) connecting $\varrho_a$ with $\varrho_b$

In our special case the fidelity is given by the formula of Eq.(\ref{fidelity}).
Thus in this case our conclusion is that there is a special POVM providing the optimal set for distinguishability. It is provided by the projectors onto a basis which diagonalizes $M=L_{ba}$. For each $E_i$ taken from this set the maximum of (\ref{maximum}) is attained. This maximum value coincides with the geodesic length calculated between $\varrho_a$ and $\varrho_b$ with respect to our (\ref{hyphyp}) metric.

\section{Geodesics and optimal measurements}

Let us also give the explicit form of our $M$.
In this section it is convenient to express everything in terms of the pair $(u,v)$ rather than the one $(a,b)$.
See Eqs.(\ref{elsoverzio}) and (\ref{renormalized}). Then from the calculations in the Appendix yielding our ${\mathcal U}_{ba}$ (see Eq.(\ref{opt2}) for its expression) using Eq.(\ref{baromi})
it is easy to obtain the corresponding form for an operator ${\mathcal L}_{vu}$ with its upper left block giving our  $M=L_{vu}$ and the lower right one its inverse $L_{uv}$.
For the calculation of $L_{vu}$ we have to use (\ref{back3}) for multipication ${\mathcal U}_{vu}$ with its $v$ dependent version from the left and its $u$ dependent one from the right.
These considerations then yield the formula
\begin{equation}
    M=L_{vu}=
    \frac{1}{2\cos D}\left[\left(1+\frac{C_v}{C_u}\right)I +\slashed{v}-\frac{C_v}{C_u}\slashed{u}\right]
\label{emm}
\end{equation}
where $\cos D(v,u)=\sqrt{F(v,u)}$ and for the formula for the fidelity see (\ref{fidelity}).
Hence the operator implementing the best measurement is featuring the geodesic length between $\varrho_u$ and $\varrho_v$ and the ratios of the quantities ${\mathcal C}_v$ and ${\mathcal C}_u$ which are the simple generalizations of the concurrences familiar from the one-qubit case.  
Since these are in turn related to the (\ref{freeenergy}) von-Neumann entropy they are serving as entanglement measures
for the entangled states purifying our density matrices.
One can also check that ${\rm Tr}(M\varrho_u)=\sqrt{F(v,u)}=\vert\langle b\vert a\rangle\vert$.

One can also immediately see that $M$ is Hermitian. In order to check that it is also positive we calculate its $2^{N-1}$-fold degenerate eigenvalues: $M_{\pm}$:
\begin{equation}
    M_{\pm}=\frac{1}{{\mathcal C}_u}\frac{{\mathcal C}_{ar}}{\cos D}\left( 1 \pm 
    \sqrt{1-\left(\frac{{\mathcal C}_{ge}}{{\mathcal C}_{ar}}\right)^2\cos^2D}\right)
\end{equation}
Here ${\mathcal C}_{ar}=({\mathcal C}_v+{\mathcal C}_u)/2$ is the arithmetic and 
${\mathcal C}_{ge}=\sqrt{{\mathcal C}_v{\mathcal C}_u}$
is the geometric mean and.
Since $\frac{{\mathcal C}_{ge}}{{\mathcal C}_{ar}}\leq 1$
the discriminant cannot be negative and the eigenvalues are indeed positive.
Notice also that the inverse of $M$ is obtained by simply exchanging $u$ and $v$.
This shows that $\varrho_v\sharp \varrho_u^{-1}$ and 
$\varrho_u\sharp \varrho_v^{-1}$
are inverses of each other. This is in accord with the intuition that parallel transporting on a path and in both direction has no effect on the state.

Let us now recall the form of a parallel lift of a geodesic between the states $\varrho_u$ and $\varrho_v$  having a geodesic distance $D:=D(v,u)$.
This is just a curve $W(t)$ in the total space of our bundle projecting to the geodesic $\varrho(t)$, which is connecting the two points $W_v=W(D)$ and $W_u=W(0)$ with the tangent vectors being horizontal for all $t$. 
Since $W_v=MW_u$ the explicit form is\cite{Uhlmann3,Ericsson}
\begin{equation}
    W(t)=\left(I\cos t +(M-I\cos D)\frac{\sin t}{\sin D}\right)W_u:=X(t)W_u
\end{equation}
Since $M$ is hermitian the projected curve is
$\varrho(t)=X(t)\varrho_uX(t)$.

Now in our special case one can use Eq.(\ref{emm}) to obtain the explicit form
\begin{equation}
    X(t)=\frac{1}{\sin 2D}\left[ \sin(2D-t)I+\sin t\left(\slashed{v}+\frac{C_v}{C_u}(I-\slashed{u}\right)\right]
\end{equation}
yielding the geodesic
\begin{equation}
    \varrho_{vu}(t)=\frac{1}{2^N}\left(I+\slashed{m}_{vu}(t)\right),\qquad
    {\bf m}_{vu}(t)=\frac{\sin(2t){\bf v}+\sin(2D-2t){\bf u}}{\sin(2D)}
\label{geodesic1}
\end{equation}
Now a calculation shows that
\begin{equation}
\vert\vert{\bf m}_{vu}(t)\vert\vert^2=1-\frac{[
C_v\sin(2t) +C_u\sin (2D-2t)]^2}{\sin^2(2D)}
\label{geodesic3}
\end{equation}
where $\sin^2(2D)=1-({\bf vu}+C_vC_u)^2$.
One can then see that the geodesic is confined to ${\mathbb B}^{2N}$ for all parameter values $t$ except
for the values $t^{\ast}$ satisfying the equation 
$
C_v\sin(2t^{\ast}) +C_u\sin (2D-2t^{\ast})=0$
in which case we will have a $2^{N-1}$-fold degenerate zero eigenvalue.
For the $N=1$ case this corresponds to the boundary of the Bloch ball, i.e. pure states where the entanglement entropy is zero. For $N\geq 2$ the boundary value of the entropy is $S=(N-1)\log 2$.
The corresponding special value $t^{\ast}$ satisfies the equation
\begin{equation}
    \cot 2t^{\ast} =\frac{{\bf vu}+C_vC_u-\frac{C_v}{C_u}}
    {\sqrt{1-({\bf vu}+C_vC_u)^2} }
\end{equation}

Now we are at the stage of clarifying the physical meaning of the basic quantity of this paper namely the special $2N$-qubit entangled state $\vert \psi _v\rangle$ which by virtue of (\ref{sokqubit}) corresponds to the special (\ref{purify}) purification $W_v$ of our density matrix $\varrho_v$. (For the change of coordinates in these quantities see Section 4.) 
This state can be created as follows.

Prepare the left-right system in the state $\vert\varphi_0\rangle$ which is of the form
\begin{equation}
   \vert \varphi_0\rangle=\frac{1}{\sqrt{2^N}}\sum_{I=0}^{2^N-1}\vert I\rangle_{\ell}\otimes\vert I\rangle_r=\bigotimes_{i=1}^{N}\frac{1}{\sqrt{2}}(\vert 0\rangle_{\ell}\otimes\vert 0\rangle_r +
    \vert 1\rangle_{\ell}\otimes\vert 1\rangle_r)
\label{vacuum}
\end{equation}
i.e. it is the $N$ fold tensor product of the EPR state $(\vert 00\rangle +\vert 11\rangle)/\sqrt{2}$.
We will call this $2N$ qubit state the {\it vacuum state}. Clearly this vacuum state is the completely entangled state of $N$ pairs of qubits. Note that $\vert\varphi_0\rangle$ is just the TFD state reinterpretation of the state underlying the construction of a type $II_1$ algebra of Murray and von Neumann. For this construction and its TFD reinterpretation see Eq. (3.11) of Ref.\cite{Witten}.
For the left system the left marginal of this vacuum state is
$\varrho_0=I/2^N$ which is just our density matrix with ${\bf u}=0$.

Now using (\ref{emm}) we perform an optimal measurement $M_v=L_{v0}$ on $\vert\varphi_0\rangle$ which yields the state 
$(L_{v0}\otimes I)\vert\varphi_0\rangle$.
The choice ${\bf u}=0$, $C_u=1$ then shows that
the resulting state will be just $\vert\varphi_v\rangle$. This is the entangled state arising from parallel transporting the vacuum along the geodesic connecting $\varrho_0$ and $\varrho_v$. From Eq.(\ref{geodesic1}) we see that this geodesic is just a straight line connecting the center of ${\mathbb B}^{2N}$ with the point ${\bf v}$.
Finally act with the unitary $U_v$ according to the pattern of Eq.(\ref{good2}) to produce the state $\vert\psi_v\rangle$.

Now if by using Eq.(\ref{coord12}) we switch from the coordinates
$v\equiv (\chi,{\bf v})\in S^1\times{\mathbb B}^{2N}$ coordinates
$(\chi,\beta,{\bf n})$ to the ones $(\tau, r,{\bf n})$ where $(\tau,{\bf r})$ are reminiscent of (Euclidean) time and space coordinates 
one can write
\begin{equation}
    \vert\varphi_v\rangle\leftrightarrow
    W(0,{\bf r}),\qquad \vert\psi_v\rangle\leftrightarrow W(\tau,{\bf r})
\end{equation}
Notice that both of them are arising via the action of suitable operators on the vacuum
\begin{equation}
    \vert\varphi_v\rangle =(M_v\otimes I)\vert\varphi_0\rangle,
    \qquad
    \vert\psi_v\rangle =(U_vM_v\otimes I)\vert\varphi_0\rangle
\label{fieldtheory}
\end{equation}
Since $L_{v0}=\sqrt{2^N}\varrho_v^{1/2}$
and $W(\tau,{\bf r})$ is normal then $U_vM_v=M_vU_v$ corresponds to the polar decomposition of $W(\tau,{\bf r})$.

\section{Interference}
In this paper we have investigated such evolutions of the entangled left$+$right system which gives rise to a sequence of optimal measurements when studied from the perspective of the left system. We have also learned that on the other hand from the perspective of the right system this evolution is a quantum computation consisting of a sequence of unitary quantum gates.
Now we would like to illustrate how the result of this computation  
can be detected by interference experiments performed on the {\it right} system. 
In order to see this first we have to display how the density matrix of the right system changes before it is subjected to an observation due to the interference experiment.

First notice that according to  Eqs. (\ref{firstex})-(\ref{megallapodas}) for the starting entangled state $\vert\varphi_a\rangle$ for the reduced density matrices of the left and right systems respectively we have the marginals
\begin{equation}
\varrho_a^{(0)}=\bar{\omega}_a^{(0)}
\label{equal}
\end{equation}
where both of these quantities have the (\ref{root}) explicit form.
After the (\ref{next})
transformation featuring unitaries, 
$\varrho_a^{(0)}$
is not changing but 
$\bar{\omega}_a^{(0)}$ is changing according to
\begin{equation}
    \bar{\omega}_a=\left(U_a^{-1}U_{ab}U_{bc}U_{ca}\right)\bar{\omega}_a^{(0)}\left(U_a^{-1}U_{ab}U_{bc}U_{ca}\right)^{\ast}={\mathcal U}\varrho_a^{(0)}{\mathcal U}^{\ast}
    \label{change1}
\end{equation}
where we used (\ref{equal}) and for the overall unitary incorporating the anholonomy effect we introduced the notation ${\mathcal U}$.

In Ref.\cite{Sjöquist1} 
the authors investigate the problem of detecting mixed state geometric phases in interferometry.
They consider
a conventional Mach-Zehnder interferometer in which
the beam pair spans a two dimensional Hilbert space
which here we regard as an extra qubit attached to our {\it right} $N$-qubit subsystem.
 In this extra-qubit picture the basis vectors of this qubit can be mapped to
wave packets that represent a particle having the possibility of moving only in two given directions defined
by the geometry of the interferometer. Then one can represent mirrors, beam splitters, and relative 
phase shifts by $2\times 2$ unitary operators.
For their explicit forms see Ref.\cite{Sjöquist1}.
The only gate which we recall is their phase gate of the form
\begin{equation}
    U(\tilde{\chi})=\begin{pmatrix}e^{i\tilde{\chi}}&0\\0&1\end{pmatrix}
\end{equation}
which produces a phase shift
$\tilde{\chi}$ in one of the paths of the split beams of the interferometer.
This is the phase that can be observed in the output signal of the interferometer.

\begin{figure}[!h]
    \centering\includegraphics[width=0.5\textwidth]{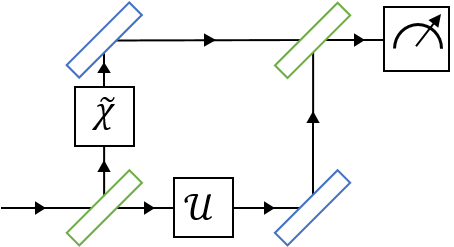}
    \caption{Illustration of the interferometric arrangement of Ref.\cite{Sjöquist1} adapted to our thermofield double setup.
    The Mach-Zehnder interferometer is regarded as an "observer" represented by a particle featuring an extra qubit which is coupled to the right system. The blue rectangles refer to mirrors and the green ones to half-silvered mirrors providing two possible paths for the particle inside the interferometer.
Then the basis vectors of the extra qubit can be regarded as the
wave packets that represent the particle having the possibility of moving only in the two given directions defined
by the geometry of the interferometer.
On the other hand the $N$-qubits of the right system are regarded as internal degrees of freedom of the particles showing up in the interferometer.
In one of the interferometric paths we apply the unitary ${\mathcal U}$ arising from Uhlmann's anholonomy associated to a closed curve. On the other hand in the other path
we apply the phase gate with phase shift $\tilde{\chi}$.
The couping is described by Eq.(\ref{SJ}). The interference pattern is recorded at the upper right corner of the setup.}
    \label{fig:inter}
\end{figure}

In this interferometric picture we can regard our right $N$-qubit subsystem as a system representing some internal degrees of freedom of the particles in the interferometer.
Then our change of the right marginal of Eq.(\ref{change1}) can be viewed as a change occurring within this internal space of the particles subject to the measurement implemented by the interferometric setup. 
One can couple this internal space to our extra qubit as follows\cite{Sjöquist1}
\begin{equation}
    {\bf 1}\otimes {\mathcal U}\otimes \begin{pmatrix}0&0\\0&1\end{pmatrix}
    +{\bf 1}\otimes {\bf 1}\otimes 
    \begin{pmatrix}e^{i\tilde{\chi}}&0\\0&0
    \end{pmatrix}
\label{SJ}
\end{equation}
Here the first two tensor product factors refer to our usual left and right subsystems and the rightmost tensor product factor corresponds to the extra qubit. 
Eq.(\ref{SJ}) shows that in one of the interferometric paths we apply ${\mathcal U}$, on the other hand in the other path 
we apply the phase gate.

Let the incoming state of this new interferometric setup is
\begin{equation}
    {\bf 1}\otimes\omega^{(0)}_a\otimes \vert 0\rangle\langle 0\vert
\label{subject}
\end{equation}
Then suppose that this state is
split coherently by a beam
splitter and then recombined at a second beam splitter after being reflected by two mirrors. We then arrange that the manipulations coming from optimal measurements in the left system make their presence in the form of the Uhlmann anholonomy ${\mathcal U}$ in the right one. Then we couple according to Eq.(\ref{subject}) this $\mathcal U$ 
to the path between the first beam splitter and the corresponding mirror. 
Then the result of Ref.\cite{Sjöquist1} shows that the output density matrix arising after (\ref{subject}) has been subjected to the interferometer
will be producing a phase shift $\Phi =\arg{\rm Tr}(\overline{\mathcal U}\omega^{(0)}_a)=-
\arg{\rm Tr}({\mathcal U}\varrho^{(0)}_a)$
 in the output intensity of the form
\begin{equation}
    {\mathcal I}\simeq 1+\vert {\rm Tr}(\overline{\mathcal U}\omega^{(0)}_a)\vert
    \cos(\tilde{\chi}-\arg{\rm Tr}(\overline{\mathcal U}\omega^{(0)}_a))
    =
    1+\vert {\rm Tr}({\mathcal U}\varrho^{(0)}_a)\vert
    \cos(\tilde{\chi}+\arg{\rm Tr}({\mathcal U}\varrho^{(0)}_a))
\label{intensity}
\end{equation} 
The quantity $\nu:=\vert {\rm Tr}({\mathcal U}\varrho^{(0)}_a)\vert$ is called the visibility of the interference pattern.

In the following we calculate the phase shift and visibility for different examples of our model. In order to do this recall Eq.(\ref{rewritehyp}) to write
\begin{equation}
    {\mathcal U}=U_a^{-1}{\mathcal R}^{-1}
\label{anholo}
\end{equation}
where for the explicit forms of $U_a$ and $\mathcal R$ see Eqs. (\ref{ua}) and (\ref{forgatasos}).

\subsection{Detecting the basic purification by interference}

Let us first consider the very simple special case where the 
anholonomy is arising purely from a parallel transport from the center of ${\mathbb B}^{2N}$ to a point ${\bf a}$ along the geodesic that is a straight line segment.
Clearly after switching labels this is just the case discussed at the end of Section.11. Namely we have to relabel Eq.(\ref{fieldtheory}) as $\vert\psi_a\rangle =U_aM_a\otimes I\vert\varphi_0\rangle$.
Here the optimal measurement is followed by the unitary, and taken together these are implementing $W_a$.  If we write out the $a\equiv(\chi,{\bf a})$ or $(\tau,{\bf r})$
coordinate dependence explicitly we can regard this implementation as the creation of the "field" $W(\tau,{\bf r})$ from the vacuum.

Now in this simple case ${\mathcal R} =I$ and in (\ref{anholo}) we have only $U_a$ hence the interference pattern is coming from the formula
\begin{equation}
    {\rm Tr}(\varrho_a^{(0)}U_a^{-1})=
    \sqrt{\frac{1+(1-C_u)^2\tan^2(\chi/2)}{1+\frac{1-C_u}{1+C_u}\tan^2(\chi/2)}}e^{i(\Phi-\chi)},\qquad \tan\Phi =\frac{C_u\sin\chi}{2-C_u+C_u\cos\chi}
\label{allati}
\end{equation}
Hence the manipulation  registered as an optimal measurement on the left is interpreted on the right hand side by an observer conducting the interference experiment as the appearance of a total phase shift $\Phi-\chi$. The visibility is given by the square root factor showing up in Eq.(\ref{allati}).

Notice that the (\ref{intensity}) formula for the intensity features the phase shift $\tilde{\chi}$ whose gate is a characteristic object of the interferometer. Since the total phase appearing in the argument of the cosine function is of the form $\tilde{\chi}-\chi+\Phi$ one can obtain the maximal intensity by adjusting $\tilde{\chi}$ accordingly. When $C_u$ is small, corresponding to small entanglement encoded into $W_a$, then $\Phi$ will be small and then by  $\tan\Phi=\Phi+\dots $ we have maximal intensity for
$\tilde{\chi}=\chi-\frac{1}{2}C_u\sin\chi+\dots$. 
Hence, for small entanglement for maximal interference, the value of this $\tilde{\chi}$ essentially should coincide with the value of $\chi$ up to a fluctuating correction depending on the entanglement.
Note that in this respect according to the instanton interpretation of the Uhlmann connection the parameter $\chi$ that can be measured in this way is playing the role of some sort of Euclidean time coordinate.
It is amusing to see that this "time" coordinate is directly related to a phase shift in the interference pattern.
Moreover, if we have a look at the unitary (\ref{ua}) appearing in the polar decomposition one can see that $\chi$ shows up in two different ways. On the one hand it is present in a factorized $U(1)$ phase factor, and on the other hand, it is featuring the $SU(2^N)$ matrix. This clarifies why $\chi$ is also connected to the other parameter $\beta$ that according to (\ref{freeenergy}) labels the strength of entanglement via $C_u=1/\cosh\beta$.  It is also worth noticing that since in this setup $\cos 2D=C_u$
the entanglement is directly linked to the geodesic length of our line segment as measured by the Bures metric.

\subsection{Geodesic triangles}

Here we would like to evaluate  interference terms for anholonomies based on Eqs.(\ref{intensity}) and (\ref{anholo}) for geodesic triangles.  First we write
${\mathcal R}^{-1}$ based on the result of Eq.(\ref{mester}) in a form convenient for our purposes.
We note that
\begin{equation}
1+2{\bf pq}+p^2q^2=(1+a^2)^2\frac{1+2{\bf bc}+b^2c^2}{(1+2{\bf ab}+a^2b^2)(1+2{\bf ac}+a^2c^2)}    
\end{equation}
and
\begin{equation}
    1+2{\bf ab}+a^2b^2=\frac{4F_{uv}}{(1+C_u)(1+C_v)}
\end{equation}
where for the (\ref{fidelity}) fidelity we have introduced the shorthand notation $F_{uv}\equiv F(u,v)$.
Then using this and Eq.(\ref{atteres}) one can show that
\begin{equation}
    \frac{1}{2}\frac{[\slashed{p},\slashed{q}]}{\sqrt{1+2{\bf pq}+p^2q^2}}
    =\frac{[\slashed{w},\slashed{v}]}{8\sqrt{F_{uv}F_{vw}F_{wu}}}+\dots
\end{equation}
where the dots refer to terms of the form $[\slashed{u}, \slashed{x} ]$ where $\slashed{x}={\bf x}\mathbf{\Gamma}$
is some matrix that we will not need at this stage.
We also recall that
\begin{equation}
\frac{(1+{\bf pq})I}{\sqrt{1+2{\bf pq}+p^2q^2}}=\cos\frac{\delta}{2}I=\frac{F_{uv}+F_{vw}+F_{wu}-1}{2\sqrt{F_{uv}F_{vw}F_{wu}}}I
\label{cd}
\end{equation}
a formula familiar from Eq.(\ref{crossszorzatos}).
Then the form of ${\mathcal R}^{-1}$ is
\begin{equation}
    {\mathcal R}^{-1}=\frac{(F_{uv}+F_{vw}+F_{wu}-1)I-\frac{1}{4}[\slashed{w},\slashed{v}]+\dots }{2\sqrt{F_{uv}F_{vw}F_{wu}}}
\end{equation}
On the other hand we also have
\begin{equation}
    \varrho_a^{(0)}U_a^{-1}=e^{-i\chi}\left( A(C_u,\chi)I+B(C_u,\chi)\slashed{u}\right)
\end{equation}
with
\begin{equation}
    A=\frac{1}{2^N}\sqrt{\frac{1+C_u}{1+C_u\cos\chi}}\left(1+iC_ue^{i\chi/2}\sin(\chi/2)\right),\qquad
B=\frac{1}{2^N}\frac{1+C_ue^{i\chi/2}\cos(\chi/2)}{\sqrt{(1+C_u)(1+C_u\cos\chi})}
\nonumber
\end{equation}

What we need is an expression for ${\rm Tr}(\varrho_a^{(0)}U_a^{-1}{\mathcal R}^{-1})$.
For the calculation of this quantity the $\dots$ terms do not give any contribution since ${\rm Tr}(\slashed{u}[\slashed{u},\slashed{x}])=0$. 
Furthermore the ${\rm Tr}(\slashed{u}[\slashed{w},\slashed{v}])=u^jw^kv^l{\rm Tr}(\Gamma_j\Gamma_k\Gamma_l)$ term also gives zero unless $N=1$.
Indeed, for $N=1$ we have ${\rm Tr}(\sigma_j\sigma_k\sigma_l)=2i\varepsilon_{jkl}$ but for $N\geq 2$ this term is zero.
Hence the result for $N\geq 2$ is
\begin{equation}
    {\rm Tr}(\varrho_a^{(0)}U_a^{-1}{\mathcal R}^{-1})=
    \cos{\frac{\delta}{2}}\sqrt{\frac{1+(1-C_u)^2\tan^2(\chi/2)}{1+\frac{1-C_u}{1+C_u}\tan^2(\chi/2)}}e^{i(\Phi-\chi)}
\label{allati2}
\end{equation}
where $\Phi$ is again given by Eq.(\ref{allati}).
This shows that there is no sign of the anholonomy in the phase shift, the anholonomy makes its presence only in the visibility with modification coming from the $\cos{\frac{\delta}{2}}$ term which is of the (\ref{cd}) form.
.

The $N=1$ case is special for geodesic triangles.
In order to see this just notice that
${\rm Tr}(\slashed{u}[\slashed{v},\slashed{w}])=-4i{\bf u}({\bf v}\times{\bf w})=-4iV$ hence this 
term is not vanishing. Here $V$ is the volume of the parallelepiped formed by the three vectors ${\bf u},{\bf v},{\bf w}$.
In this case we have the explicit formula
\begin{equation}
{\rm Tr}(\varrho_a^{(0)}U_a^{-1}{\mathcal R}^{-1})=
\frac{e^{-i\chi}}{2\sqrt{F_{uv}F_{vw}F_{wu}}}\left[A(F_{uv}+F_{vw}+F_{wu}-1)+\frac{i}{2}BV\right]
\label{inter1}
\end{equation}
For $\chi=0$ this formula boils down to Eq.(34) of Ref.\cite{LPThomas}.

Clearly for $N>1$ geodesic triangles always give rise to the much simpler formula of Eq.
(\ref{allati2}). In order to also obtain terms containing "V-terms" coming from volumes one should consider higher dimensional volumes associated with geodesic $(2N+1)$-gons.
For example for $N=2$ the relevant terms will be arising from ones that contain ${\rm Tr}(\Gamma_j\Gamma_k\Gamma_l\Gamma_m\Gamma_n)\simeq \varepsilon_{jklmn}$.
Such a term is associated with geodesic pentagons. 
An alternative possibility for obtaining such V-terms is to consider the combined anholonomy effect of {\it two} geodesic triangles say $abc$ and $ade$. If we first traverse $abc$ then we get again back to the starting point $a$. Then next we traverse $ade$. Now instead of inserting in (\ref{allati2}) merely the inverse of ${\mathcal R}(a,b,c)$ of Eq.(\ref{mester}) 
one should insert the inverse of the product 
${\mathcal R}(a,b,c){\mathcal R}(a,d,e)$.

One can also realize that one can go for realizing merely a particular set of gates implementing universality.
We note in this respect that one possibly needs only to consider an interesting and useful set of quantum gates. 
To such an example we turn to next.

\subsection{Universal gates from Uhlmann's anholonomy}

Here we would like to show that the anholonomy arising from geodesic triangles can be used to perform universal quantum computations. For this purpose we investigate the $iSWAP$ gate
\begin{equation}
    iSWAP= U_{iS}=   \begin{pmatrix}
        1 & 0 & 0 & 0 \\
        0 & 0 & i & 0 \\
        0 & i & 0 & 0 \\
        0 & 0 & 0 & 1
    \end{pmatrix}=e^{i\frac{\pi}{4}(\sigma_1 \otimes \sigma_1+\sigma_2 \otimes \sigma_2)}=e^{i\frac{\pi}{4}\sigma_1 \otimes \sigma_1}e^{i\frac{\pi}{4}\sigma_2 \otimes \sigma_2}
    \label{iSWAP}
\end{equation}
 this can be shown by straightforward calculation. For splitting the exponent into two we also used that $[\sigma_i\otimes\sigma_i,\sigma_j\otimes\sigma_j]=0$. Our goal is now to find a geodesic triangle that generates these two terms.
 To this end let us consider the following simple three point configuration.  Start from the origin so $\textbf{a}=0$ and let us choose the two other points as $\textbf{b}=b{\bf e}_i$ and $\textbf{c}=c{\bf e}_j$. Then based on Eqs.(\ref{inversion1}) and (\ref{inversion2}) we can write $\textbf{p}={\textbf c}$,
$\textbf{q}={\textbf b}$,
         $\slashed{p}=c\Gamma_i$ and
         $\slashed{q}=b\Gamma_j$.
 Now the anholonomy derived in Eq. (\ref{mester}) takes the following simple form
 \begin{equation}
         {\mathcal R}({\bf a},{\bf b},{\bf c})=\frac{I+\frac{1}{2}bc[\Gamma_i,\Gamma_j]}{\sqrt{1+b^2c^2}}
 \end{equation}

Let us now fix $N=3$ and use the Jordan-Wigner type realization of the gamma matrices as given by Eq.(\ref{rekur2}).
Then one can check that
\begin{equation}
[\Gamma_0,\Gamma_3]=-2i\sigma_2\otimes \sigma_2\otimes {\bf 1},\qquad
[\Gamma_1,\Gamma_2]=2i\sigma_1\otimes \sigma_1\otimes {\bf 1}
\end{equation}
\begin{equation}
[\Gamma_2,\Gamma_5]=-2i{\bf 1}\otimes\sigma_2\otimes \sigma_2,\qquad
[\Gamma_3,\Gamma_4]=2i{\bf 1}\otimes\sigma_1\otimes \sigma_1
\end{equation}
Then one has
\begin{equation}
{\mathcal R}({\bf 0},b{\bf e}_2,c{\bf e}_1)
=\left(\frac{{\bf 1}\otimes{\bf 1}i+ibc\sigma_1\otimes\sigma_1}{\sqrt{1+(bc)^2}}\right)\otimes {\bf 1},\quad
{\mathcal R}({\bf 0},b{\bf e}_0,c{\bf e}_3)
=\left(\frac{{\bf 1}\otimes{\bf 1}+ibc\sigma_2\otimes\sigma_2}{\sqrt{1+(bc)^2}}\right)\otimes {\bf 1}
\nonumber
\end{equation}
\begin{equation}
{\mathcal R}({\bf 0},b{\bf e}_4,c{\bf e}_3)
={\bf 1}\otimes\left(\frac{{\bf 1}\otimes{\bf 1}+ibc\sigma_1\otimes\sigma_1}{\sqrt{1+(bc)^2}}\right),\quad
{\mathcal R}({\bf 0},b{\bf e}_2,c{\bf e}_5)
={\bf 1}\otimes\left(\frac{{\bf 1}\otimes{\bf 1}+ibc\sigma_2\otimes\sigma_2}{\sqrt{1+(bc)^2}}\right)
\nonumber
\end{equation}

From this one can see that with the choice $\sin(\delta/2)=bc/\sqrt{1+b^2c^2}$, one can reproduce the iSWAP either in the first two or the last two qubits with the choice $\delta=\pi/2$. However, this leads to an intresting problem, namely that one should have $\vert\textbf{b}|=|\textbf{c}|=1$, which means that we need to go to the edge of our Bloch-ball and then include non-invertible states. 
We can avoid this problem if we split the exponent into two and use the angle $\delta=\pi/4$ instead, namely
\begin{equation}
U_{iS}=e^{i\varphi\sigma_1\otimes\sigma_1}e^{i\varphi\sigma_1\otimes\sigma_1}e^{i\varphi\sigma_2\otimes\sigma_2}e^{i\varphi\sigma_2\otimes\sigma_2}
\end{equation}
with $\varphi=\delta/2=\pi/8$.
Then instead of two geodesic triangles one should use four ones to arrive at the $iSWAP$ gate.

This approach shows that there is not a single anholonomy generating an $iSWAP$ gate, but continuum many.
In particular one can choose the symmetric special case when $b=c=\tanh{\beta/2}$. Then with the choice $\delta=\pi/4$ we get 
$\beta=\log(\sqrt{2}+1)$.
This means that $\vert{\bf b}\vert=\vert{\bf c}\vert=\sqrt{2}-1=0,414\dots$. All of these triangles are consisting of two straight segments from the origin to $\bf b$ and $\bf c$, and a geodesic segment of the form (\ref{geodesic3}) stretching between $\bf b$ and $\bf c$.

Hence the Uhlmann anholonomy with these parameter choices is capable of producing an iSWAP gate in the first and second or in the second and third qubits.
The most important thing about the $iSWAP$  gate is that taken together with the one qubit operations it realizes a complete set of unitary quantum gates\cite{iSWAP1}, \cite{iSWAP2} needed for computational universality.

\section{Conclusions and comments}

In this paper a simple parametrized family of quantum systems, consisting of two entangled subsystems, dubbed left and right ones, both of them featuring $N$ qubits has been considered in the thermofield double formalism.
Changing the external parameters classically via prescribing certain paths in the space of parameters results in a changing set of entangled states.
We assumed that due to the change in parameters the system evolves in a purely geometric manner based on the parallel transport condition due to Uhlmann.
The physical meaning of this condition has its roots in the basic phenomenon of quantum interference adapted to entangled states.
In our special case this means that two entangled states of $2N$ qubits are parallel if and only if under the variation of local unitary transformations acting on {\it one of} the $N$ qubit subsystems the interference is extremal.

We explored the different interpretations of this evolution relative to observers either coupled to  the left or to the right subsystems. Clearly the Uhlmann condition breaks the symmetry between left and right by regarding one of the two possible sets of local unitary operations as gauge degrees of freedom. This space of local unitaries comprises the manifold where extremization is to be taken into account.
In this paper, we have chosen gauging the right subsystem. Then we have shown that on the left a single discretized step of this geometric evolution manifests itself by a local operation reminiscent of a {\it non-unitary} filtering measurement or equivalently as an optimal measurement for distinguishing between the two marginals featuring the step in the statistical sense. 
On the other hand on the right the basic evolutionary steps are organized into a sequence of {\it unitary} operations of a holonomic quantum computation. 
We
calculated the Uhlmann connection governing the parallel transport for our model and we have found that the associated gauge field is related to higher dimensional instantons. 
Then we evaluated the
anholonomy of the connection for geodesic triangles with geodesic segments where geodesics are defined with respect to the
Bures metric.
Then we analyzed the explicit form of
the local filtering (optimal) measurements showing up on the left side and we have also explicitly given the mathematical form of the optimal measurements.
We have also pointed out that by conducting an
interference experiment on the right side
one can observe the physical effects of the resulting anholonomic quantum computation.
We demonstrated this by calculating explicit examples for phase shifts and visibility patterns arising in such experiments. Finally a sequence of geodesic triangles producing the iSWAP gate via anholonomy needed for computational universality was presented.

In the introduction we have emphasized that in this paper we are not exploring the explicit form of the dynamics driving the full ${\mathcal S}+{\mathcal E}$ system. We were simply assuming that the interaction between system and environment is somehow implements Uhlmann's parallelity via a dynamics giving rise to our geometric type of evolution. Hence our paper was about exploring the consequences of a possible geometric evolution rather than about finding the culprit for such an evolution.

Let us however, finally give here some interesting hints on the possible physical origin of such an evolution.
Recall first Eqs. (\ref{crucial}) and (\ref{crucial2}).
These expressions describe the interpretations of our evolution from the right and left perspectives in two formalisms of very different kind.
These are the formalisms of entangled states with local manipulations on left and right, and the one of "block spinors". 
For the correspondence between these formalisms see Eq.(\ref{blockspin}).

The reader then can discover that
the right hand side of Eq.(\ref{blockspin}) establishes the framework of this paper for studying an anholonomy effect (Uhlmann's phase) for {\it mixed states}.
On the other hand the left hand side does the same for an anholonomy effect for subspaces of {\it pure states}.
Indeed, the block spinor formalism describes evolving $n=2^N$ dimensional subspaces within the framework of the non-adiabatic non-Abelian geometric phase\cite{Anandan}. In this case the non-Abelian gauge-structure then can be realized as a degree of freedom in choosing basis states in the degenerate subspace. This freedom manifests itself in the freedom of {\it right} multiplication by $U_a\in U(n)$
of the canonical section in the corresponding Stiefel bundle (with base space the Grassmannian of $n$ planes through the origin of the $2n$ dimensional vector space). 
Apart from this right action there is a corresponding {\it left} action as well. It is the action of the group $U(2n)$ corresponding to the time evolution operator usually expressed as a time ordered exponential generated by {\it some} Hamiltonian. 

It is easy to identify the generator of this time evolution. Let us define
\begin{equation}
    h(X):=X_0\Gamma_0+{\bf X}\cdot{\mathbf \Gamma }+X_{2N+2}\Gamma_{2N+2}
\label{velocity}
\end{equation}
where the coordinates are the ones familiar from Section 7.
Then the unitary matrix of Eq.(\ref{first1}), having the alternative (\ref{uuu}) appearance
 needed for the (\ref{pur1}) characterization of our entangled state,
has the property $U^{\ast}(X)h(X)U(X)=\Gamma_{2N+2}$ i.e. it diagonalizes $h(X)$.
Define now
\begin{equation}
{\mathcal 
    H}(X(t)):=\frac{1}{4i}[h(X(t),h(\dot{X}(t))]
\label{SchHam}
\end{equation}
Then the
block spinor type description is just the discretized one of evolving $n$-dimensional subspaces
under the Schrödinger equation with the time dependent Hamiltonian of Eq.(\ref{SchHam}) under the restriction of $X_0=0$.
Indeed, generalizing Eq.(4.26) of Ref.\cite{Levayquat} it is easy to show that
\begin{equation}
    {\mathbb T}e^{-i\int_0^T{\mathcal H}(t)dt}\vert a\rangle=\vert a\rangle {\mathbb P}e^{-i\int_{\mathcal C}{\mathcal A}}
\label{spin1}
\end{equation}
where ${\mathbb T}$ and ${\mathbb P}$ denote time and path ordering respectively and ${\mathcal C}$ is a closed curve traversed in time $T$. Note also that ${\mathcal A}$ formally coincides with Uhlmann's connection of Eq.(\ref{Uconnection2}) arising as the restriction of the higher dimensional instanton connections via the constraint $X_0=0$, see Eq.(\ref{UhlmannInstant}).
Notice also that the evolution equation of (\ref{spin1}) is {\it} {\it not} describing the non-Abelian generalization\cite{Wilczek} of Berry's Phase. It is rather the non-Abelian generalization of the Aharonov-Anandan phase\cite{Anandan}.
Indeed, during evolution $\vert a(t)\rangle$ is not an instantaneous eigenstate of ${\mathcal H}(t)$ the generator of Schrödinger evolution, it is an eigenstate of $h(t)$.
Moreover, as has been observed in Ref.\cite{Karlhede} under the evolution generated by ${\mathcal H}(t)$ the dynamic phase
is exactly zero 
due to $\langle a(t)\vert {\mathcal H}(t)\vert a(t)\rangle=0$
.
For the non-Abelian case generally the dynamic and the geometric phase parts of the evolution cannot be separated\cite{Anandan}. However, now the dynamical contribution is zero
hence, fulfilling the basic assumption of this paper, the evolution in (\ref{spin1}) is indeed purely geometric.

In closing this paper we would like to draw the readers attention to the fact that there is an important physical situation where the discretized version of the time evolution operator on the left hand side of Eq.(\ref{spin1}) appears.
As familiar from Eq.(\ref{crucial2}) this evolution can be written as a sequence of rank $n$ projectors i.e. filtering measurements of the form $\vert a\rangle\langle a\vert /\vert\langle a\vert a\rangle\vert$.
Now it is a well-known result that the Dirac propagator 
$\Delta (x,y)$ 
 of a spinning particle propagating in Euclidean space can be represented
 in terms of Polyakov's spin factors\cite{Polyakov,Korchemsky} which can be cast to the form of a sequence of such projectors.
 This approach is based on Strominger's analysis of the propagator in loop space\cite{Strominger}, i.e. in the the space of paths connecting $x$ and $y$.
In this approach the propagator is written as the path ordered product along the particle orbit $x(s)$ with $s\in[0,T]$ subject to the boundary conditions $x(0)=x$ and $y=x(T)$. Now after identifying the normalized velocity vector $\dot{x}$ of the particle with our $X$ of Eq.(\ref{velocity}) 
we realize
that
 the propagator encapsulates a cumulative effect of succesive Thomas rotations\cite{Nowak} along the particle path. Notice that these rotations are precisely the same objects studied in this paper but in a different context.
We are planning to study this interesting analogy in a forthcoming publication.

\section{Acknowledgement}

Acknowledgment: This work was supported by the HUN-REN Hungarian Research Network through the HUN-REN-BME-BCE Quantum Technology Research Group.

\section{Appendices}

\subsection{Gamma matrix representations}

The gamma matrices featuring (\ref{anti}) are special matrices in $2N+1$ dimensions. We  denote them as
$\Gamma^{(2N+1)}_j$ with $ j=0,1,2,\dots 2N$.
We use two different realizations in the text.

One of them is defined  recursively as follows.
With the
notation $\hat{j}=1,2,\dots 2N-1$ and $\Gamma^{(1)}_0=I^{(1)}=1$ the recursive procedure is given by
\begin{equation}
\Gamma_0^{(2N+1)}=\sigma_1\otimes I^{(2N-1)}.\qquad
\Gamma_{\hat{j}}^{(2N+1)}=\sigma_2\otimes \Gamma_{\hat{j}-1}^{(2N-1)},\qquad
\Gamma_{2N}^{(2N+1)}=\sigma_3\otimes I^{(2N-1)}
\label{rekurgamma}
\end{equation}
Here the $I^{(2N+1)}$ are the $2^N\times 2^N$ identity matrices.
If we use the notation 
\begin{equation}
I^{(3)}:= {\bf 1},\qquad I^{(2N+1)}:=I,\qquad \Gamma_j:=\Gamma^{(2N+1)}
\end{equation} 
then for $N=1$ we have
\begin{equation}
\Gamma_0=\sigma_1,\qquad \Gamma_1=\sigma_2,\qquad \Gamma_2=\sigma_3    
\end{equation}
for $N=2$
\begin{equation}
\Gamma_0=\sigma_1\otimes {\bf 1},\quad \Gamma_1=\sigma_2\otimes\sigma_1,\quad \Gamma_2=\sigma_2\otimes \sigma_2,\quad \Gamma_3=\sigma_2\otimes \sigma_3,\quad \Gamma_4=\sigma_3\otimes {\bf 1}    
\label{rek2}
\end{equation}
for $N=3$
\begin{equation}
\Gamma_0=\sigma_1\otimes {\bf 1}\otimes {\bf 1},\quad \Gamma_1=\sigma_2\otimes\sigma_1\otimes {\bf 1},\quad \Gamma_2=\sigma_2\otimes \sigma_2\otimes\sigma_1,\quad\Gamma_3=\sigma_2\otimes \sigma_2\otimes\sigma_2
\nonumber\end{equation}
\begin{equation}
\Gamma_4=\sigma_2\otimes\sigma_2 \otimes\sigma_3,
\quad \Gamma_5=\sigma_2\otimes \sigma_3\otimes {\bf 1},\quad
\Gamma_6=\sigma_3\otimes {\bf 1}\otimes {\bf 1}
\end{equation}
etc. This means that each recursive step is featuring the gamma matrices of the previous recursive step. Namely, the old one is showing up in the tensor  product factors of the new one, to the right of the leftmost Pauli matrix $\sigma_2$.

The other realization is the one which should be familiar from Jordan-Wigner transformation.
Neglecting in this case the tensor product symbols and the $2\times 2$ identity matrix factors moreover, by  denoting the Pauli matrices as $(\sigma_1,\sigma_2,\sigma_3):=(X,Y,Z)$ this $N$-qubit operator representation is summarized as
\begin{equation}
\Gamma_{2I-2}:=Z_1Z_2\dots Z_{I-1}X_I,\quad
\Gamma_{2I-1}:=Z_1Z_2\dots Z_{I-1}Y_I,\quad \Gamma_{2N}:=Z_1Z_2\cdots Z_{N}
\label{rekur2}
\end{equation}
where $I=1,2,\dots N$. Here the notation $Z_{I-1}$ etc. means that the $I-1$th tensor product slot is occupied by the operator $Z$. 
Note that $\Gamma_{2N}=(-i)^N\Gamma_0\Gamma_1\cdots \Gamma_{2N-1}$ is the parity operator.
Clearly in this picture the first realization given by Eq.(\ref{rekurgamma}) can be obtained from (\ref{rekur2})
after exchanging $Z$ and $Y$ and permuting the labels accordingly.
Note that for such a Jordan-Wigner type realization for $N=2$ from Eq.(\ref{rekur2}) we get the explicit forms 
\begin{equation}
\Gamma_0={\sigma_1}\otimes {\bf 1},\qquad 
\Gamma_1={\sigma_2}\otimes{\bf 1},\qquad
\Gamma_2={\sigma_3}\otimes{\sigma}_1,\qquad
\Gamma_3={\sigma_3}\otimes{\sigma}_2,\qquad \Gamma_4={\sigma}_3\otimes{\sigma}_3 
\label{kellenifog}
\end{equation}
which is to be compared with Eq.(\ref{rek2}).

\subsection{On the purification of our density matrix}

Since $W$ is normal i.e. $W^{\ast}W=WW^{\ast}$ 
it can be written in the polar decomposed form
\begin{equation}
W={\varrho}^{1/2}U=
U{\varrho}^{1/2}
, \qquad U^{\dagger}=U^{-1}   
\label{polar11}
\end{equation}
Indeed, one can check that
$W$ of (\ref{purify}) can be given the alternative form
\begin{equation}W=\left(\frac{\Omega}{2^{N+1}\cosh\beta}\right)^{1/2}\left[e^{i\chi}I+e^{-\beta H({\bf n})}\right]
\label{purify2}
\end{equation}
This shows that with
\begin{equation}
\varrho^{1/2}=\left(\frac{1}{2^N\cosh\beta}\right)^{1/2}
e^{-\frac{1}{2}\beta H}>0
\label{para}
\end{equation}
and the unitary $U$ 
\begin{equation}
U=e^{\frac{i}{2}\chi}\left(\frac{\Omega}{2}\right)^{1/2}\left[e^{-\frac{i}{2}\chi}e^{-\frac{1}{2}\beta H} +e^{\frac{i}{2}\chi}
e^{\frac{1}{2}\beta H}\right]\in U(1)\times SU(2^N)
\label{uniegy}
\end{equation}
the polar decomposition holds.
An alternative form for the unitary $U$ to be used later is
\begin{equation}
    U=e^{\frac{i}{2}\chi}(2\Omega)^{1/2}\left[(\cosh\beta/2\cos\chi/2)I-i(\sinh\beta/2\sin\chi/2)\slashed{\bf n}\right] 
\label{sjöquist}
\end{equation} 
The matrix $\varrho^{1/2}$
regarded as the canonical purification of $\varrho$ is not depending on $\chi$.

We also note that after diagonalizing $H({\bf n})$
\begin{equation}
H({\bf n})=\slashed{\bf n}=S\begin{pmatrix}{\bf 1}&0\\0&-{\bf 1}\end{pmatrix}S^{\ast}
\end{equation}
where $\bf 1$ is the $2^{N-1}\times 2^{N-1}$ unit matrix one can also write a diagonal form for the purification $W$ of Eq.(\ref{purify}) as 
\begin{equation}
    SWS^{\ast}=e^{\frac{i}{2}\chi}\frac{1}{\sqrt{Z_N}}
    \begin{pmatrix}
e^{-\beta/2}{\bf 1}&0\\0&e^{\beta/2}{\bf 1}        
    \end{pmatrix}
\begin{pmatrix}
\sqrt{\frac{\cosh(\zeta_+/2)}{\cosh(\zeta_-/2)}}{\bf 1}&0\\        
0&\sqrt{\frac{\cosh(\zeta_-/2)}{\cosh(\zeta_+/2)}}{\bf 1}    
    \end{pmatrix}
\end{equation}
where
\begin{equation}
 \zeta_{\pm}:=\beta\pm i\chi   
\end{equation}
A comparison of this form with (\ref{uniegy}) clearly shows that the matrix $U$ is an element of the group $SU(2^N)$.
Notice that thanks to the gamma matrices showing up in our considerations, we should rather regard $U\in {\rm Spin}(2N+1)\subset SU(2^N)$.

\subsection{Calculating the Uhlmann connection}

We use the split $I=(\alpha\bar{\alpha})$ with $\alpha,\bar{\alpha} =1,2,\dots,2^{N-1}$. Here the two sets of $2^{N-1}$-fold degenerate eigensubspaces with eigenvalues $+1$ and $-1$ of the operator $\hat{\slashed{\bf u}}$ are spanned by the eigenvectors $\vert \alpha\rangle$ and $\vert\bar{\alpha}\rangle$ respectively.
Notice that
\begin{equation}
    P_{\pm}:=\frac{1}{2}(I\pm 
    \hat{\slashed{\bf u}}),\qquad \hat{\bf u}:={\bf u}/\vert {\bf u}\vert\in S^{2N}
\label{unit}
\end{equation}
are the spectral projectors of rank $2^{N-1}$. 
We have $P_\pm^2=P_\pm$ and $P_\pm P_\mp=0$ moreover $P_++P_-=\mathbb{I}$. Then from the spectral resolution $\rho=P_+(1+|\mathbf{u}|)/2^N+P_-(1-|\mathbf{u}|)/2^N$.
These projectors have the properties 
\begin{equation}
    P_+\vert \alpha\rangle=\vert \alpha\rangle,\qquad
    P_+\vert\bar{\alpha}\rangle=0,\qquad
    P_-\vert \bar{\alpha}\rangle=\vert \bar{\alpha}\rangle,\qquad
    P_-\vert \alpha\rangle=0
\end{equation}

Let us first calculate the eigenvalue part of the sum. If $I\in \{\alpha\}$ and $J\in \{\beta\}$ $\mathcal{A}=0$ because the eigenvalues are identical, the same is true if $I\in\{\bar{\alpha}\}$ and $J\in \{\bar{\beta}\}$ $\mathcal{A}=0$. The cases that remain are the ones with $I\in \{\alpha\},J\in \{\bar{\beta}\}$ 
and $I\in \{\bar{\alpha}\},J\in \{ \beta\}$. For both of them we get
\begin{equation}
    \frac{(\sqrt{\lambda_I}-\sqrt{\lambda_J})^2}{\lambda_I+\lambda_J}=1-\sqrt{1-\vert{\bf u}\vert^2}
\end{equation}
a result which is independent from all indices.

Now we have
\begin{equation}
{\mathcal A}=(\sqrt{1-\vert{\bf u}\vert^2}-1)\left[ \sum_{\alpha,\bar{\beta}}\vert \alpha\rangle\langle \alpha\vert d\vert \bar{\beta}\rangle\langle \bar{\beta}\vert +
\sum_{\bar{\alpha},\beta}\vert \bar{\alpha}\rangle\langle \bar{\alpha}\vert d\vert \beta\rangle\langle \beta\vert
\right]
\end{equation}
Now we have $P_-\vert\bar{\beta}\rangle=\vert\bar{\beta}\rangle$ and
$P_+\vert \beta\rangle=\vert \beta\rangle$ then
$dP_-\vert \bar{\beta}\rangle +P_- d\vert\bar{\beta}\rangle
=d\vert \bar{\beta}\rangle$
. 
Moreover, we have
$\sum_{\alpha} \vert {\alpha}\rangle\langle    \alpha\vert =P_+$
and
$\sum_{\bar{\alpha}} \vert \bar{\alpha}\rangle\langle \bar{\alpha} \vert =P_-$.
hence
\begin{equation}
\sum_{\alpha,\bar{\beta}}\vert \alpha\rangle\langle \alpha\vert d\vert \bar{\beta}\rangle\langle \bar{\beta}\vert =\sum_{\bar{\beta}}P_+
d\vert \bar{\beta}\rangle\langle\bar{\beta}\vert =\sum_{\bar{\beta}}\left(P_+dP_-\vert \bar{\beta}\rangle\langle\bar{\beta}\vert +P_+P_- d\vert\bar{\beta}\rangle\langle\bar{\beta}\vert\right)=P_+dP_-P_-    
\nonumber
\end{equation}
and
\begin{equation}
\sum_{\bar{\alpha},\beta}\vert \bar{\alpha}\rangle\langle \bar{\alpha}\vert d\vert \beta\rangle\langle \beta\vert =P_-dP_+P_+    
\end{equation}
This yields
\begin{equation}
    {\mathcal A}=
    (\sqrt{1-\vert{\bf u}\vert^2}-1)\left[P_+dP_-P_- +
    P_-dP_+P_+
    \right]
\end{equation}
Since $P_+P_-=0$ we have $dP_-P_++P_-dP_+=0$ etc. Using this and
$P_{\pm}^2=P_{\pm}$ gives the more compact form
\begin{equation}
    {\mathcal A}=
    (1-\sqrt{1-\vert{\bf u}\vert^2})\left[dP_+P_- +
    dP_-P_+
    \right]
\end{equation}
Now $dP_{\pm}=\pm\frac{1}{2}d\hat{\bf u}{\boldsymbol {\Gamma}}$
and $P_{\pm}=\frac{1}{2}(I\pm\hat{\bf u}{\boldsymbol{\Gamma}})$
hence
\begin{equation}
    {\mathcal A}=
    \frac{1}{2}(\sqrt{1-\vert{\bf u}\vert^2}-1)(d\hat{\bf u}{\boldsymbol{\Gamma}})
    (\hat{\bf u}{\boldsymbol{\Gamma}})
\end{equation}
Now
\begin{equation}
    (d\hat{\bf u}{\boldsymbol{\Gamma}})
    (\hat{\bf u}{\boldsymbol{\Gamma}})=
    \frac{1}{2}[\Gamma_j,\Gamma_k]d\hat{u}^j\hat{u}^k+
    \frac{1}{2}\{\Gamma_j,\Gamma_k\}d\hat{u}^j\hat{u}^k=-[\Gamma_k,\Gamma_j]\hat{u}^kd\hat{u}^j=-[\hat{\bf u}{\boldsymbol{\Gamma}},
    d\hat{\bf u}{\boldsymbol{\Gamma}}
    ]
\end{equation}
since $\frac{1}{2}\{\Gamma_j,\Gamma_k\}d\hat{u}^j\hat{u}^k=d\hat{\bf u}\hat{\bf u}=0$ due to 
(\ref{anti}) and (\ref{unit}).
Hence
\begin{equation}
    {\mathcal A}=
    \frac{1}{4}(1-\sqrt{1-\vert{\bf u}\vert^2})
[{\bf n}{\boldsymbol{\Gamma}},
    d{\bf n}{\boldsymbol{\Gamma}}]
\label{Uconnectionap}
\end{equation}
If we again use the $\tanh\beta =|\mathbf{u}|^2$ parametrization we can write that
    $1-\sqrt{1-\vert{\bf u}\vert^2}=1-\text{sech}\beta$
In these new coordinates
the pullback of the Uhlmann connection is
\begin{equation}
    {\mathcal A}=
\frac{1}{4}(1-\text{sech}(\beta))
    [\hat{\slashed{u}},
    d\hat{\slashed{u}}]=
\frac{1}{4}\left(1-\sqrt{\frac{(1-r^2)^2+(1+\tau^2)^2-1}
{(1+r^2)^2+(1+\tau^2)^2-1}}\right)
[\hat{\slashed{u}},
    d\hat{\slashed{u}}]
\label{Uconnection2ap}
\end{equation}

In order to relate Uhlmann's connection to the higher dimensional monopole gauge-fields of Ref.\cite{Zalanek} we proceed as follows. First note that in this context according to Eq.(\ref{szokasos}) we had used the definition $\hat{\bf u}=-{\bf n}$. 
Then we calculate the anti-Hermitian parts
\begin{equation}
{\rm Im}(XdX^{\dagger})=i(X_0d{\bf X}-{\bf X}dX_0){\boldsymbol{\Gamma}}+\frac{1}{2}[{\bf X}{\boldsymbol{\Gamma}},d{\bf X}{\boldsymbol{\Gamma}}]    
\end{equation}
\begin{equation}
{\rm Im}(X^{\dagger}dX)=-i(X_0d{\bf X}-{\bf X}dX_0){\boldsymbol{\Gamma}}+\frac{1}{2}[{\bf X}{\boldsymbol{\Gamma}},d{\bf X}{\boldsymbol{\Gamma}}]    
\end{equation}
And then express the result in terms of the $\tau,{\bf r}$ coordinates defined by (\ref{angles1})-(\ref{sproj}).
Then
\begin{equation}
X_0d{\bf X}-{\bf X}dX_0=\frac{4}{(1+\tau^2+r^2)^2}(\tau d{\bf r}-{\bf r}d\tau),\qquad \frac{1}{2(1+X_{2N+2})}=\frac{1+\tau^2+r^2}{4}    
\end{equation}
\begin{equation}
\frac{1}{2}[{\bf X}{\boldsymbol{\Gamma}},d{\bf X}{\boldsymbol{\Gamma}}]=\frac{2r^2}{(1+\tau^2+r^2)^2}    
[{\bf n}{\boldsymbol{\Gamma}},d{\bf n}{\boldsymbol{\Gamma}}]
\end{equation}
With these results we get
\begin{equation}
    A_{\pm}=\pm\frac{i}{1+\tau^2+r^2}({\bf r}d\tau-\tau d{\bf r}){\boldsymbol{\Gamma}}+
    \frac{r^2}{2(1+\tau^2+r^2)}    
[{\bf n}{\boldsymbol{\Gamma}},d{\bf n}{\boldsymbol{\Gamma}}]
\label{instantonform1}
\end{equation}
With the restriction $\tau=0$ we obtain
\begin{equation}
A_{\pm}\vert_{\tau=0}=    
\frac{r^2}{2(1+r^2)}    
[\slashed{n},d\slashed{n}]
\end{equation}
This coincides with the $\tau=0$ restriction of $\mathcal A$ of Eq.(\ref{Uconnection2ap}) as claimed in the main text.
A similar calculation using (\ref{explicitparallel}) for the parallel translation operators gives the result of Eq.(\ref{Upara}).

\subsection{Coordinates}

For the clarification of the meaning of Uhlmann's connection we needed a coordinate transformations. We review them here. 
First define ${\bf r}:=r{\mathbf n}$ and let
\begin{equation}
    X_0=\frac{\sin{\chi}}{\cosh\beta}=
    \frac{2\tau}{1+\tau^2+r^2},\qquad
    {\bf X}=\tanh{\beta}{\bf n}=
    \frac{2{\bf r}}{1+\tau^2+r^2}
    \label{angles1}
\end{equation}
\begin{equation}
X_{2N+2}=\frac{\cos{\chi}}{\cosh\beta}=\frac{1-\tau^2-r^2}{1+\tau^2+r^2}
\label{angles2}
\end{equation}

where $j=1,2,\dots 2N+1$ and ${\bf n}\in S^{2N}$. Now ${\bf X}\in {\mathbb R}^{2N+1}$ and one can see that

\begin{equation}
    X_0^2+\vert{\bf X}\vert^2+X_{2N+2}^2=1
\end{equation}
i.e. a fixed $X\in {\mathbb R}^{2N+3}$ with coordinates $X_{\mu},\quad \mu=0,1,2,\dots 2N+1,2N+2$
defines a point on the sphere $S^{2N+2}$.
One can also verify that
\begin{equation}
    \tau=\frac{X_0}{1+X_{2N+2}},\qquad
    {\bf r}:=r{\bf n}=\frac{\bf X}{1+X_{2N+2}}
    \label{sproj}
\end{equation}
In the main text via the instanton analogy we suggested regarding $\tau$ and $X_0$ as Wick rotated (Euclideanized) time coordinates.
Let us then depict Euclidean time to proceed from bottom to top. Hence the $X_0$ or $\tau=0$  i.e. "time" axis is vertical.

In this picture taking $X_0=\tau=0$ corresponds to taking a Euclideanized "static slice" (ESS). Topologically the ESS
is a sphere $S^{2N+1}$ that corresponds to the "equator" of $S^{2N+2}$. It is like the equator of the two-sphere that is formally obtained by plugging in $N=0$ and which is now  indeed a circle $S^1$.
Similarly the analogue of the "Greenwich meridian" corresponds to the locus of points with $X_{2N+2}=0$. This is another $S^{2N+1}$. The intersection of the equator with the Greenwich meridian is the locus $X_0=X_{2N+2}=0$ and then the resulting submanifold  yields an $S^{2N}$. For illustration for $N=0$ this intersection consists of just two points since $S^0\simeq{\mathbb Z}_2$.

This shows that the coordinates 
$\tau$ and ${\bf r}:=r{\bf n}$ are stereographic images of the points of the sphere $S^{2N+2}$, projected from the point $X=(0,\dots,0,-1)$ to the "plane" containing the "Greenwich meridian":  ${\mathbb R}^{2N+2}$.

The first physically interesting case is $N=1$. In this case we have two $S^3$s intersecting in an $S^2$ which is just the boundary of the Bloch sphere.
In this case $\varrho$ will be degenerate on this $S^2$ since in this case it will be a rank $2^{N-1}=1$ projector. 
Indeed, for general $N\geq 1$ from Eq.(\ref{asfollows}) we see that
\begin{equation}
{\rm Det W}\simeq(X_{2N+2}+iX_0)^{2^{N-1}}
\end{equation}
hence our $\varrho=WW^{\ast}$ degenerates here.
Since in this paper we restrict attention to full rank density matrices the physics of this $S^{2N}$ is not considered here.

For the right "hemisphere" we have  
$X_{2N+2}>0$ for the left one $X_{2N+2}<0$.
We also see that in order to have the static slice one should either have $\chi=0$ or $\chi=\pi$. Then from Eq.(\ref{angles2}) one can see that in the first case we are on the right  part of the "equator" and in the second we are on the left. In the first case we have $r=\tanh(\beta/2)\leq 1$ and in the second $r=\coth(\beta/2)\geq1$.

We obtain an important insight into the geometry of our model if we employ the coordinate $C:=1/\cosh\beta$ familiar from Eq.(\ref{meghat}).
Since the (\ref{freeenergy}) entanglement entropy is a function of $\beta$ for a measure of the entanglement between left and right one can alternatively use $C$. For $N=1$ this is just the concurrence.
Now in terms of $C$ one can recast Eq.(\ref{angles1})-(\ref{angles2})
to the form
\begin{equation}
X_{2N+2}=C\cos\chi,\qquad X_0=C\sin\chi,\qquad {\bf X}=\sqrt{1-C^2}{\bf n},\qquad 0<C\leq 1   
\label{angles3}
\end{equation}
Hence the pair $(X_{2N+2},X_0)$ parametrizes a collection of circles with radius $0<r_1\leq 1$ with $r_1=C$. Likewise ${\bf X}$ parametrizes ${\bf}$ $2N$ spheres with radius $0\leq r_2<1$ where the radii are related by $r_1^2+r_2^2=1$.
Then this (\ref{angles3}) way of writing up the Cartesian coordinates of $S^{2N+2}$ shows that this sphere is sliced up by to submanifolds of the $S_{r_1}^1\times S_{r_2}^{2N}$ form characterized by different entanglement.
It is easy to show that the line element $d\ell^2$ on this 
$S^{2N+2}$ is of the form
\begin{equation}
d\ell^2=\left(\frac{1}{\cosh\beta}\right)^2\left[d\chi^2+d\beta^2+\sinh^2\beta d\vert{\bf n}\vert ^2\right]
\end{equation}
This shows that these submanifolds patch together conformally to yield the conformal equivalence $S^{2N+2}\simeq {\mathbb B}^{2N}\times S^1$.
Here recall that according to Eq.(\ref{Ballset}) the interior of the generalized Bloch ball i.e. the manifold ${\mathbb B}^{2N}$ is $2N+1$ dimensional.
Notice that up to a factor of $4$ the part of this line element which corresponds to ${\mathbb B}^{2N}$ is (according to (\ref{hyphyp})) the Bures metric.

The upshot is that the parameter space for the special purification of Eq.(\ref{purify}) defining our parametrized family of entangled states is conformally the manifold ${\mathbb B}^{2N}\times S^1$. However, if we look at the relationship between the coordinates $\chi$ and the ones $(\beta,{\bf n})$
more closely then we realize that they boil down to the coordinates of $S^{2N+2}$, an object rather identified as a twisted product  than a Cartesian product manifold.
This twisted structure is then  manifested e.g. in the complicated anholonomy expressions for the interference visibility of Eq.(\ref{allati}).

\subsection{Calculating the anholonomy}

In this Appendix we would like to calculate
$U_{ij}$ of Eq.(\ref{final})
the basic building block of the Uhlmann anholonomy.
Before calculating this we recall that  
\begin{equation}
    L(v)\gamma^{\nu}L^{-1}(v)={\Lambda_{\mu}}^{\nu}(v)\gamma^{\mu},\qquad \mu,\nu=0,1,2,\dots,2N+1
    \label{use2}
\end{equation}
where
\begin{equation}
    {\Lambda_0}^0(v)=\cosh\beta_v,\qquad
    {\Lambda_0}^k(v)={\Lambda_k}^0(v)=-\sinh\beta_v{\hat v}^k
\end{equation}
\begin{equation}
    {\Lambda_k}^j(v)={\delta_k}^l+(\cosh\beta_v -1){\hat v}_k{\hat v}^j
\end{equation}
which can be checked by a straightforward calculation. Notice that the matrix elements are the standard boost elements in the vector representation.

Noticing that $L^2(u)\gamma^0=\cosh\beta_u\gamma^0-\sinh\beta_u(\hat{\bf u}\boldsymbol{\gamma})$ and using (\ref{use2}) one can then prove that
\begin{equation}
    L(v)L^2(u)L(v)=-L(v)\left[L^2(u)\gamma^0\right]L^{-1}(v)\gamma^0=L^2(w)
\label{naezkell}
\end{equation}
where
\begin{equation}
    L^2(w)=\cosh\beta_w I+\sinh\beta_w{(\hat{\bf w}}\boldsymbol{\gamma})\gamma^0
\label{takingroot}
\end{equation}
with
\begin{equation}    \cosh\beta_w=\cosh\beta_u\cosh\beta_v+\sinh{\beta}_u\sinh\beta_v(\hat{\bf u}\hat{\bf v})   
\label{kifejez}
\end{equation}
and
\begin{equation}
\sinh\beta_w\hat{\bf w}=\cosh\beta_u\sinh\beta_v\hat{\bf v} +\sinh\beta_u(\cosh\beta_v -1)(\hat{\bf u}\hat{\bf v})\hat{\bf v}+\sinh\beta_u\hat{\bf u}   
\end{equation}
\noindent
Notice that for $\hat{\bf u}=\hat{\bf v}$ (collinear boosts) we get
$\cosh\beta_w=\cosh(\beta_u+\beta_v)$ and
$\sinh\beta_w\hat{\bf w}=(\cosh\beta_u\sinh\beta_v+\sinh\beta_u\cosh\beta_v)\hat{\bf v}$
which yields the additivity of the rapidities: $\beta_w=\beta_u+\beta_v$ and $\hat{\bf w}=\hat{\bf v}$.

Now clearly one can take the square root of (\ref{takingroot}) to get 
\begin{equation}
    L(w)=\cosh\frac{\beta_w}{2}I+\sinh\frac{\beta_w}{2}
    (\hat{\bf w}\boldsymbol{\gamma})\gamma^0
\label{elw}
\end{equation}
Using these results we obtain
\begin{equation}
L(v)L^2(u)L(v)+I=2\cosh\frac{\beta_w}{2} L(w)  
\label{idesuss}
\end{equation}
moreover, from (\ref{naezkell}) we also realize that 
\begin{equation}
    (L(v)L(u)L(u)L(v))^{1/2}=L(w)
\label{square}
\end{equation}
Then
\begin{equation}
    L^{-1}(u)L^{-1}(v)(L(v)L(u)L(u)L(v))^{1/2}=
    L^{-1}(u)L^{-1}(v)L(w)=\frac{L^{-1}(u)L^{-1}(v)+L(u)L(v)}{2\cosh\frac{\beta_w}{2}}
\nonumber
\end{equation}
Now one can use Eq.(\ref{back2}) to discover that
\begin{equation}
    \frac{L^{-1}(u)L^{-1}(v)+L(u)L(v)}{2\cosh\frac{\beta_w}{2}}=\begin{pmatrix}\varrho_u^{-1/2}\varrho_v^{-1/2}\left(\varrho_v^{1/2}\varrho_u\varrho_v^{1/2}\right)^{1/2}&0\\0&
\varrho_u^{1/2}\varrho_v^{1/2}\left(\varrho_v^{-1/2}\varrho_u^{-1}\varrho_v^{-1/2}\right)^{1/2}\end{pmatrix}
\nonumber
\end{equation}
hence in the upper left corner of this matrix we discover $U_{uv}$ our basic building block.
Let us denote the matrix on the right hand side by ${\mathcal U}_{uv}$
Using the explicit form of the matrices $L(u)$ and $L(v)$
for the matrix ${\mathcal U}_{uv}$ one obtains the formula of Eq.(\ref{mesterke}).

\bigskip
We know that $U_{uv}$ is unitary.  More precisely it shows up as an $SU(2^N)$ matrix as a diagonal block of ${\mathcal U}_{uv}$ which is an element of the group $SU(2^{N+1})$.
One can check this as follows.
First notice that $L^{\ast}(u)=L(u)$. Then we have 
\begin{equation}
    {\mathcal U}_{uv}^{\ast}=
\frac{L^{-1}(v)L^{-1}(u)+L(v)L(u)}{2\cosh\frac{\beta_w}{2}}
\label{opt}
\end{equation}
Then
\begin{equation}
{\mathcal U}^{\ast}_{uv}{\mathcal U}_{uv}=
\frac{L^{-1}(v)L^{-1}(u)+L(v)L(u)}{2\cosh\frac{\beta_w}{2}}
\frac{L^{-1}(u)L^{-1}(v)+L(u)L(v)}{2\cosh\frac{\beta_w}{2}}
\label{opt2}
\end{equation}
Hence by virtue of (\ref{naezkell})
\begin{equation}
U^{\ast}_{uv}U_{uv}
=\frac{2I+[L^2(w)+L^{-2}(w)]}{4\cosh^2\frac{\beta_w}{2}}=
\frac{2(1+\cosh\beta_w)I}{4\cosh^2\frac{\beta_w}{2}}=I
\label{lenyeg2}
\end{equation}
Let us now consider the quantity
\begin{equation}
    iB:=\frac{2\hat{u}^j\hat{v}^k\Sigma_{jk}}{\sqrt{1-(\hat{\bf u}\hat{\bf v})^2}}
\end{equation}
Then one can check that $B^{\ast}=B$.
Now looking at Eq.(\ref{lenyeg}) and 
(\ref{lenyeg2}) one can see that we also have the property $B^2=I$ then ${\mathcal U}_{uv}$ can be written in the form
\begin{equation}
    {\mathcal U}_{uv}=e^{i(\delta/2)B}
\end{equation}
Now since ${\rm Tr}B=0$ this formula shows that ${\rm Det}({\mathcal U}_{uv})=1$, hence 
${\mathcal U}_{uv}\in SU(2^{N+1})$.

\subsection{Calculation of the Bures metric}

From Eq. (10) and (F.17) of Ref.\cite{china} it is known that for a full rank density matrix the Bures metric can be calculated from the formula

\begin{equation}
    g_{jk}=  \frac{1}{2}{Re}\sum_{IJ} \frac{(\partial_j \rho)_{IJ}(\partial_k\rho)_{JI}}{\lambda_I+\lambda_J} 
    \label{buresmetric}
\end{equation}
where the matrix elements are taken with respect to the eigenbasis of $\varrho$ and $\partial_j$ denotes derivatives with respect to the parameters $u^j$.
Explicitly
\begin{equation}
    g_{jk}=\frac{1}{2^{2N+1}} {\rm Re}\sum_{IJ}\frac{(\Gamma_j)_{IJ}(\Gamma_k)_{JI}}{\lambda_I+\lambda_J}
    \label{bugamma}
\end{equation}

For the calculation of the metric we use the (\ref{unit}) projectors and their set of eigenvectors corresponding to the degenerate eigensubspaces.
First note that due to
 $(\Gamma_j \hat{u}^j)\Gamma_k=-\Gamma_k(\Gamma_j\hat{u}^j)+ 2\hat{u}_j I$
 one has $P_{\pm}\Gamma_j=\Gamma_j P_{\mp}\pm\hat{u}_jI$ then 
 \begin{equation}
     \langle \alpha\vert \Gamma_j\vert \beta\rangle=
     \langle \alpha\vert P_+\Gamma_jP_+\vert \beta\rangle
     =\hat{u}_j\delta_{\alpha\beta},\qquad
\langle \bar{\alpha}\vert \Gamma_j\vert \bar{\beta}\rangle=
     \langle \bar{\alpha}\vert P_-\Gamma_jP_-\vert \bar{\beta}\rangle
     =-\hat{u}_j\delta_{\bar{\alpha}\bar{\beta}}  
 \nonumber
 \end{equation}
Now we have
\begin{equation}
    \begin{gathered}
        g_{jk}=\frac{1}{2^{2N+1}}\sum_{IJ}^{2^N} \frac{(\Gamma_j)_{IJ}(\Gamma_k)_{JI}}{\lambda_I+\lambda_J}=\frac{1}{2^{2N+1}}\left[ 2^{N-1}\sum_{\alpha\beta}^{2^{N-1}} \frac{(\Gamma_j)_{\alpha\beta}(\Gamma_k)_{\beta\alpha}}{1+|\mathbf{u}|}+ 2^{N-1}\sum_{\bar{\alpha}\bar{\beta}}^{2^{N-1}} \frac{(\Gamma_j)_{\bar{\alpha}\bar{\beta}}(\Gamma_k)_{\bar{\beta}\bar{\alpha}}}{1-|\mathbf{u}|}\right]+\\
        +\frac{1}{2^{2N+1}}\left[ 2^{N-1}\sum_{\alpha\bar{\beta}}^{2^{N-1}}(\Gamma_j)_{\alpha\bar{\beta}}(\Gamma_k)_{\bar{\beta}\alpha}+2^{N-1}\sum_{\bar{\alpha}\beta}^{2^{N-1}}(\Gamma_j)_{\bar{\alpha}\beta}(\Gamma_k)_{\beta\bar{\alpha}} \right]
    \end{gathered}
    \nonumber
\end{equation}
In order to evaluate the second term we write  $(\Gamma_j\Gamma_k)_{\alpha\alpha}=(\Gamma_j)_{\alpha\beta}(\Gamma_k)_{\beta\alpha}+(\Gamma_j)_{\alpha\bar{\beta}}(\Gamma_k)_{\bar{\beta}\alpha}$ and $(\Gamma_j\Gamma_k)_{\bar{\alpha}\bar{\alpha}}=(\Gamma_j)_{\bar{\alpha}\beta}(\Gamma_k)_{\beta\bar{\alpha}}+(\Gamma_j)_{\bar{\alpha}\bar{\beta}}(\Gamma_k)_{\bar{\beta}\bar{\alpha}}$. 
Using this we get
\begin{equation}
    \begin{gathered}
        g_{jk}=\frac{1}{2^{N+2}}\left[ \left(\frac{1}{1+|\mathbf{u}|}-1\right)\sum_{\alpha\beta}^{2^{N-1}} (\Gamma_j)_{\alpha\beta}(\Gamma_k)_{\beta\alpha}+ \left(\frac{1}{1-|\mathbf{u}|}-1\right)\sum_{\bar{\alpha}\bar{\beta}}^{2^{N-1}} (\Gamma_j)_{\bar{\alpha}\bar{\beta}}(\Gamma_k)_{\bar{\beta}\bar    {\alpha}}\right]+\\
        +\frac{1}{2^{N+2}}\left[ \sum_{\alpha}^{2^{N-1}}(\Gamma_j \Gamma_k)_{\alpha\alpha}+\sum_{{\bar{\alpha}}}^{2^{N-1}}(\Gamma_j\Gamma_k)_{\bar{\alpha}\bar{ \alpha}} \right] =\frac{1}{2^{N+2}}\left[2^N\frac{|\mathbf{u}|^2}{1-|\mathbf{u}|^2}\hat{u}_j\hat{u}_k+{\rm Tr}(\Gamma_j \Gamma_k)\right]=\\=\frac{1}{4}\left[\delta_{jk}+\frac{|\mathbf{u}|^2}{1-|\mathbf{u}|^2}\hat{u}_j\hat{u}_k\right] 
    \end{gathered}
    \nonumber
\end{equation}
Introducing the hyperbolic parametrization for the line element of the metric we obtain Eq.(\ref{hyphyp}) of the main text.

\bibliography{references}
\bibliographystyle{JHEP}




\end{document}